	\documentclass[fleqn,10pt]{wlscirep}
	\usepackage[utf8]{inputenc}
	\usepackage[T1]{fontenc}
	\usepackage{bm}
	\usepackage{caption}
	\usepackage{subcaption}
	\usepackage{parskip}
	\usepackage{amssymb}
	\usepackage{amsmath}
	\usepackage{mathtools}
	\usepackage{algorithm}
	\usepackage{graphicx}
	\usepackage{xcolor}
	\usepackage{algpseudocode}
	\usepackage{xr}
	\usepackage{afterpage}
	\usepackage{siunitx}
	\newcommand{\boldface}[1]{\boldsymbol{#1}}

	\newcommand{\bfc}{\boldface{c}}
	
	\newcommand{\bfe}{\boldface{e}}

	\newcommand{\bfk}{\boldface{k}}

	\newcommand{\bfr}{\boldface{r}}
	
	\newcommand{\bft}{\boldface{t}}

	\newcommand{\bfw}{\boldface{w}}
	\newcommand{\bfx}{\boldface{x}}
	\newcommand{\bfy}{\boldface{y}}

	\newcommand{\bfI}{\boldface{I}}

	\newcommand{\bfnull}{\boldsymbol{0}}

	\DeclareSymbolFont{matha}{OML}{txmi}{m}{it}
	\DeclareMathSymbol{\varv}{\mathord}{matha}{118}

	\newcommand{\be}{\begin{equation}}
		\newcommand{\ee}{\end{equation}}
	\newcommand{\beq}{\begin{eqnarray}}
		\newcommand{\eeq}{\end{eqnarray}}
	\newcommand{\bem}{\begin{multline}}
		\newcommand{\eem}{\end{multline}}
	\newcommand{\ba}{\begin{align}}
		\newcommand{\ea}{\end{align}}

	\newcommand{\micro}{^\mathrm{m}}

	\makeatletter
	
	\newcommand*{\addFileDependency}[1]{
	\typeout{(#1)}
	\@addtofilelist{#1}
	
	\IfFileExists{#1}{}{\typeout{No file #1.}}
	}\makeatother
	
	\newcommand*{\myexternaldocument}[1]{
	\externaldocument{#1}
	\addFileDependency{#1.tex}
	\addFileDependency{#1.aux}
	}
	\myexternaldocument{SI}
	
	\makeatletter
	
	\title{Algebraic Language Models for Inverse Design of Metamaterials via Diffusion Transformers}
	
	\author[1]{Li Zheng}
	\author[2,*]{Siddhant Kumar}
	\author[1,*]{Dennis M. Kochmann}
	
	\affil[1]{Mechanics \& Materials Laboratory, Department of Mechanical and Process Engineering, ETH Z\"{u}rich, 8092 Z\"{u}rich, Switzerland}
	\affil[2]{Department of Materials Science and Engineering, Delft University of Technology, 2628 CD Delft, Netherlands}
	\affil[*]{sid.kumar@tudelft.nl,\ dmk@ethz.ch}
	
	\begin{abstract}
	
		Generative machine learning models have revolutionized material discovery by capturing complex structure-property relationships, yet extending these approaches to the inverse design of three-dimensional metamaterials remains limited by computational complexity and underexplored design spaces due to the lack of expressive representations. Here, we present \textit{DiffuMeta}, a generative framework integrating diffusion transformers with an algebraic language representation, encoding 3D geometries as mathematical sentences. This compact, unified parameterization spans diverse topologies, enabling the direct application of transformers to structural design. DiffuMeta leverages diffusion models to generate novel shell structures with precisely targeted stress-strain responses under large deformations, accounting for buckling and contact while addressing the inherent one-to-many mapping by producing diverse solutions. Uniquely, our approach enables simultaneous control over multiple mechanical objectives, including linear and nonlinear responses beyond training domains. Experimental validation of fabricated structures further confirms the efficacy of our approach for accelerated design of metamaterials and structures with tailored properties.

	\end{abstract}
	
\begin{document}
	\flushbottom
	\maketitle
	\thispagestyle{empty}
	
	\section*{Introduction}
	
	Recent advances in additive manufacturing have enabled precise fabrication of architected (meta-)materials with unprecedented control over their properties by leveraging carefully engineered architectures across multiple length scales~\cite{surjadi2019mechanical,jiao2023mechanical}. Among the various classes of metamaterials, shell networks~\cite{han2015new,bonatti2019smoothshell} have attracted particular interest due to their exceptional stiffness-to-weight ratio~\cite{chen2020lightweight} and superior energy absorption~\cite{wang2022inverse}. Unlike beam- or plate-based lattices that exhibit stress concentrations at discrete junctions~\cite{meza2014strong,davami2015ultralight}, shell lattices utilize continuous, smoothly curved surfaces to distribute loads effectively, enabling enhanced load-bearing capability and resistance to global buckling~\cite{jung2018multiscale}. These mechanical advantages, coupled with their highly tunable topologies, position shell lattices as ideal candidates for applications including heat exchangers~\cite{iyer2022heat}, soft robotics~\cite{liu2023ultrafast}, and energy-absorbing components~\cite{abueidda2016effective}.
	
	To fully harness the design potential of architected materials, considerable efforts have focused on the inverse design problem -- identifying small-scale structures that achieve specified effective target properties. This challenge has been primarily addressed through topology optimization~\cite{xia2017recent,telgen2022topologya}, which, however, requires extensive forward finite element (FE) simulations, posing significant challenges for high-dimensional design spaces and nonlinear target properties. More recently, data-driven machine learning (ML) approaches have emerged as promising alternatives, with surrogate models providing fast property predictions~\cite{white2019multiscalea,zheng2021datadriven} and deep generative models -- such as variational autoencoders~\cite{kingma2013autoencoding} and generative adversarial networks~\cite{goodfellow2014generativea} -- demonstrating remarkable success in materials discovery~\cite{fuhr2022deep}. While significant progress has been made in leveraging ML models for the inverse design of linear material properties (such as elastic stiffness~\cite{bastek2022invertinga} or Poisson's ratio~\cite{zheng2021controllable}), extending those methods to the nonlinear regime remains challenging due to complex structure-property mappings. Recent advances have addressed this through neural network surrogates with optimization strategies~\cite{thakolkaran2025experimentinformed,deng2022inverse} and advanced generative models~\cite{bastek2023inversec,kartashov2025large}. However, existing inverse design methods have predominantly focused on optimizing for a single type of mechanical response at a time. A primary challenge lies in simultaneously achieving multiple target properties, e.g., specific linear responses under small strains, while preserving desired nonlinear behavior at large deformations. These distinct responses stem from fundamentally different deformation mechanisms that may impose conflicting requirements on the structure. The multi-objective inverse design challenge significantly increases computational complexity and intensifies the ill-posed nature of the inverse problem~\cite{kumar2020inversedesigned}.
	
	To address these challenges, diffusion models~\cite{ho2020denoising} have emerged as a promising solution for their exceptional generation quality and versatile conditioning capabilities. These models have demonstrated state-of-the-art performance across various domains, particularly in image and video synthesis~\cite{rombach2022highresolutiona}, and have recently shown remarkable potential in materials design, including crystalline materials~\cite{zeni2025generativea}, proteins~\cite{watson2023novo}, and nonlinear metamaterials~\cite{bastek2023inversec,vlassis2023denoisingb}. By learning from existing design spaces, diffusion models capture complex underlying data distributions, enabling the efficient generation of novel, coherent designs tailored to specific conditions. Despite these advances, existing works on the generative design of microstructures remain predominantly confined to two-dimensional (2D) structures with pixelated representations, limited by computational demands. While recent studies have extended to three-dimensional (3D) geometries using explicit representations such as voxels~\cite{zheng2025optimizing}, meshes~\cite{liu2023meshdiffusion}, and point clouds~\cite{luo2021diffusion}, substantial challenges persist as explicit 3D representations are inherently data-intensive. The scarcity of high-quality training data significantly hinders robust learning, leading to compromised generation fidelity~\cite{zheng2025optimizing} and constraining the practical applicability of generated designs. 
	
	These limitations become particularly acute for shell-based metamaterials. Characterized by thin, curved geometries, shell lattices exhibit pronounced geometric nonlinearities under large deformations, accompanied by complex phenomena including plasticity, buckling, and contact. Moreover, these coupled effects are highly sensitive to geometric variations, where minor structural imperfections -- such as collapsed walls, non-uniform thickness, or sharp corners -- can severely compromise the geometric integrity and trigger undesired behavior. While explicit representations provide direct descriptions, precisely capturing the curvature details of shells requires extremely high-resolution discretizations, which becomes computationally prohibitive for large-scale design exploration. In contrast, implicit representations, given by simple mathematical equations, offer an efficient and robust approach for constructing shell topologies such as, e.g., triply periodic minimal surfaces~\cite{al-ketan2019multifunctionala} and spinodoid metamaterials~\cite{kumar2020inversedesigned}. By tuning implicit equation parameters, one can precisely control the geometric features (e.g., curvature, size, symmetry, and isosurface thickness) and, consequently, tailor the mechanical properties of the resulting shell topologies, offering considerable design flexibility. Nevertheless, the relationship between implicit equations and resulting designs is a highly complex mapping -- small changes in the mathematical formulation can produce vastly different topologies and properties, while strikingly different equations may yield similar mechanical properties. This intricate mapping is a fundamental challenge for systematic design exploration. {While symbolic regression is efficient in discovering interpretable analytical relationships~\cite{udrescu2020aia}, it is primarily suited for forward modeling and requires computationally expensive outer optimization loops for inverse design.} Previous studies have primarily relied on intuitively selecting a limited set of equation templates and systematically varying numerical coefficients~\cite{wang2022inverse,jadhav2024generative}. While practical, this restricted parameteric approach narrows the design space to predefined designs (e.g., gyroid, primitive, or diamond surfaces), potentially leading to biased, suboptimal solutions, especially when multiple target properties are involved.
	
	To overcome these limitations, we introduce a mathematical language-based parameterization for shell metamaterials, enabling the systematic exploration of 3D implicit surfaces. Unlike existing works that optimize parameters within fixed equation templates, our method conceptualizes implicit equations as sequences of mathematical tokens (variables, operators, functions, and constants). This \textit{equation-as-sequence} paradigm systematically expands the design space beyond predefined classes, offering an efficient, flexible representation while retaining geometric characteristics of diverse shell topologies. To leverage this comprehensive design space, we propose \textit{DiffuMeta}, a generative modeling framework built on diffusion transformers~\cite{peebles2023scalablea} (DiT) for the inverse design of shell metamaterials with target mechanical properties. By exploiting syntactic patterns and semantic relationships in implicit equations, DiffuMeta generates diverse, physically valid shell topologies that satisfy target stress-strain responses while capturing complex physical phenomena such as buckling, plasticity, and frictional contact. Our model effectively addresses the inherent ambiguity of one-to-many mappings in inverse design tasks and, distinct from prior works targeting a single mechanical property, enables simultaneous control over multiple target properties, which may extend substantially beyond the training domain. Finally, we conduct experimental tests on 3D-printed shell structures generated by DiffuMeta to validate the model's accuracy and practical applicability.

	\section*{Results}
	\subsection*{Shell metamaterial design space}
	
	\begin{figure}[!htbp]
		\centering
		\includegraphics[width = \textwidth]{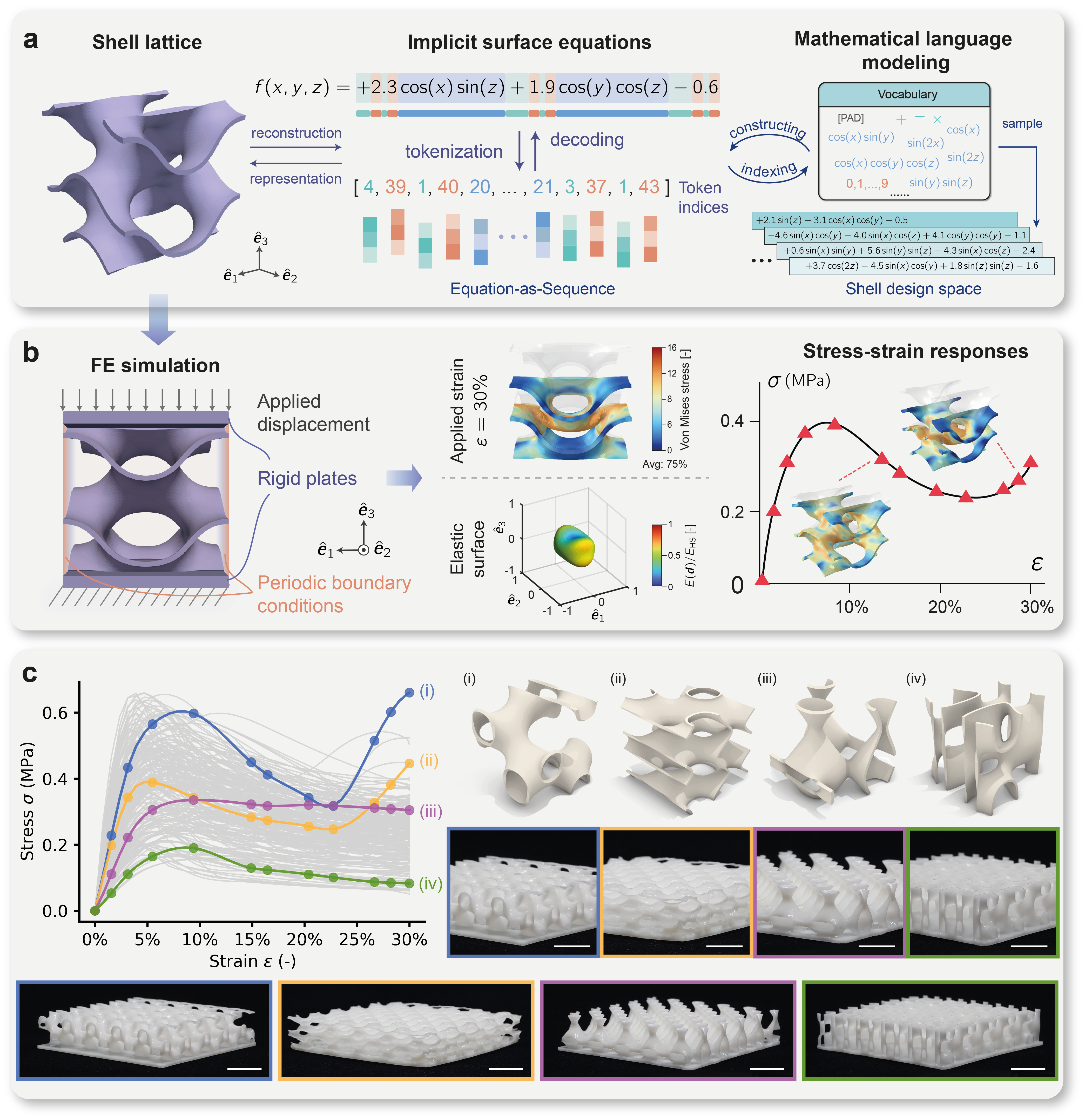}
		\captionsetup{justification=justified}
		\caption{\textbf{Overview of the shell-based metamaterial parameterization and design space generation process.} \textbf{(a)}~A shell lattice is generated from the implicit level set equation, which can be tokenized into a sequence of discrete mathematical tokens drawn from a structured vocabulary. Novel shell designs can be generated by sampling and recombining these tokens. \textbf{(b)}~To obtain the stress-strain responses, we conduct finite element (FE) simulations by imposing rigid plates on the top and the bottom with periodic boundary conditions on the lateral surfaces. A quasi-static compressive strain of up to 30\% along the $\hat\bfe_3$-direction is applied. The overall effective stress-strain response is extracted from the reaction forces. A representative deformed shell unit cell is shown at an applied strain of $30\%$. Also shown is the 3D elastic surface plot illustrating the directional effective Young's modulus (obtained via FE homogenization). \textbf{(c)}~Stress-strain responses of 200 structures randomly drawn from the dataset, and selected representative examples with distinct mechanical behaviors. Also shown are the photos of fabricated shell structures, each obtained by arranging corresponding shell unit cells in a $5 \times 5 \times 1$ array, with a total size of $50\mathrm{mm} \times 50\mathrm{mm} \times 10\mathrm{mm}$. All scale bars indicate $10\mathrm{mm}$.}
		\label{fig:methods-overview}
	\end{figure}
	
	We start by establishing a comprehensive design space for shell structures encompassing a broad spectrum of mechanical properties. To this end, we leverage periodic nodal surfaces~\cite{gandy2001nodala} (PNS), which describe smooth, continuous, and non-intersecting surfaces that are periodic in all spatial dimensions. Mathematically, periodic surfaces can be represented by the nodal surface of a sum of Fourier modes~\cite{vonschnering1991nodal}:
	\begin{equation}\label{eq:fourier-series-pns}
		\Psi(\bfr) = \sum_{\bfk}F(\bfk)\cos[2\pi\bfk\cdot\bfr- \alpha (\bfk)]= 0,
	\end{equation}
	where $\bfk$ denotes the reciprocal vectors for a given lattice, $\bfr$ defines a position vector in real space having Cartesian coordinates $\{x,y,z\}$, $\alpha(\bfk)$ represents the phase shift, and $F(\bfk)$ corresponds to the amplitude associated with~$\bfk$. For example, Gyroid surfaces can be approximated by the level set of the implicit equation
	\begin{equation}\label{eq:gyroid-eq}
		\Psi_{\text{Gyroid}}(x,y,z) = \sin(\omega x)\cos(\omega y)+\sin(\omega y)\cos(\omega z)+\sin(\omega z)\cos(\omega x)+c=0,
	\end{equation}
	where $\omega$ defines the periodicity and the constant $c$ determines the isovalue that controls the porosity of the generated surface. Such representations based on trigonometric functions provide a compact yet highly adaptable framework for generating diverse periodic shell topologies. To systematically expand the design space, we construct a library of Fourier-type basis functions $T_i(x,y,z)$ (Supplementary Table 1), including trigonometric functions (e.g., $\sin(x)$, $\cos(y)$) and their low-order multiplicative combinations (e.g., $\sin(x)\cos(y)$). Using this library, we assemble implicit surface equations by strategically combining those terms with randomly sampled coefficients. The general form of our candidate implicit equations can be represented as
	\begin{equation}
		\Psi(x,y,z) = \sum_{i \in \mathcal{I}} \alpha_{i}T_i(x,y,z) + c = 0, \quad \text{with } \mathcal{I}\subset\{1,2,\ldots,N\}, \quad 1\leq|\mathcal{I}|\leq 3,
	\end{equation}
	where $N$ is the total number of tokens in that library, $\alpha_i$ are their corresponding coefficients, and $c$ is a constant offset. To ensure physical feasibility, we verify that the generated surfaces are self-connected and periodic in all spatial dimensions, and disregard disjoint surfaces (Supplementary Section 1). {The coefficients $\alpha_i$ and constant offset $c$ are drawn from a uniformly spaced grid spanning $(-6, 6)$ in increments of~0.1, providing extensive parameter coverage while ensuring computational tractability {(see also Supplementary Table 2 for a systematic evaluation of different sampling schemes)}.}
	
	To efficiently represent this diverse design space, we develop an algebraic language-based parameterization that decomposes each equation into a sequence of discrete mathematical tokens of three types: trigonometric function groups (e.g., $\sin(x)\cos(y)$), numeric coefficients, and arithmetic operators (e.g., $+$, $-$). 
	These components define a mathematical vocabulary that enables systematic generation and analysis of implicit surfaces within a unified language-modeling framework (Fig.~\ref{fig:methods-overview}a). Any given implicit equation $\Psi(x,y,z)$ can hence be represented as a sequence $\bfw = [w_0, w_1, \ldots, w_n]$ of length $n$, each $w_i$ corresponding to the $i$-th token. For example, the Gyroid surface equation in Eq.~\ref{eq:gyroid-eq} for $\omega=1$ can be tokenized as
	\begin{equation}
		\bfw = [+,\ 1,.,0,\ \sin(y)\cos(z),\ +,\ 1,.,0,\ \sin(x)\cos(y),\ +,\ 1,.,0,\ \sin(z)\cos(x),\ \text{[PAD]},\ldots].
	\end{equation}
	Each sequence is padded to the maximum sequence length $L$ to ensure uniform dimensionality across the dataset. This \textit{equation-as-sequence} strategy provides a compact yet precise representation that preserves the mathematical integrity of implicit surfaces, with significantly reduced dimensionality compared to explicit representations. By conceptualizing implicit equations as a mathematical language, our approach captures contextual dependencies among equation components, enabling the application of natural language processing techniques to generate novel shell metamaterial designs.
	
	By systematically varying the basis functions and coefficients, we construct a rich database containing $23,534$ unique shell topologies, ensuring broad coverage of the design space with diverse mechanical behaviors (Supplementary Section 1). The wall thickness of each unit cell is scaled to maintain a constant relative density (fill fraction) of $\rho = 0.1$. To characterize their mechanical behavior, we evaluate the stress-strain responses of each design via FE analysis (Fig.~\ref{fig:methods-overview}b). The shell unit cell ($10\mathrm{mm} \times 10\mathrm{mm} \times 10\mathrm{mm}$) is positioned between two rigid plates with periodic boundary conditions along all lateral directions. Shell structures are modeled as an elastoplastic material calibrated to an in-house 3D-printing elastomer~\cite{rosa2025enhanced}, accounting for frictional contact. We apply quasi-static compressive strain up to $\varepsilon = 30\%$ through the upper plate in the $\hat\bfe_3$-direction, emulating loading scenarios for footwear cushioning~\cite{amorim2019exploring} and impact absorption~\cite{perroni-scharf2025dataefficienta}. Additionally, we compute the effective stiffness tensor through FE homogenization of a single unit cell with periodic boundary conditions. For each shell structure represented by its implicit equation $\Psi$, we collect its compressive stress response $\boldsymbol{\sigma} \in \mathbb{R}^{11}$ evaluated at 11 selected strain levels (Methods and Supplementary Section 2), as well as the corresponding homogenized stiffness tensor $\mathbb{C}\in \mathbb{R}^{6\times 6}$. This defines our dataset $\mathcal{D} = \{(\Psi^{(n)}, \boldsymbol{\sigma}^{(n)}, \mathbb{C}^{(n)}): n = 1,\dots,N\}$, comprising $N$ structure-response pairs. This yields a versatile library of periodic shell surfaces, encompassing both classical minimal surface geometries (e.g., Schwarz P and D, Gyroids) and not frequently studied morphologies, altogether exhibiting diverse stress-strain responses (Fig.~\ref{fig:methods-overview}c).
	
	\subsection*{Diffusion-based generative modeling framework}
	
	The proposed language-based parameterization provides a compact, efficient representation for diverse shell topologies. A fundamental challenge lies in the complex, many-to-many mapping between algebraic sequences and resulting shell topologies. This relationship is highly nonlinear, with seemingly distinct mathematical formulations producing similar geometries, while minor variations in coefficients may lead to drastically different topologies (Supplementary Fig. 1). Such ambiguity, coupled with the high-dimensional parameter space, creates an intricate design landscape that renders traditional optimization approaches ineffective -- motivating our use of a diffusion-based framework that learns complex structure-property relationships directly from data.
	
	Diffusion models~\cite{ho2020denoising} are probabilistic generative models that learn to reverse a gradual noising process, enabling the generation of high-quality samples from complex data distributions. While they excel in continuous domains like computer vision (using U-Net architectures~\cite{ronneberger2015unet}), their application to discrete textual data remains challenging due to the gap between continuous diffusion processes and discrete inputs~\cite{li2022diffusionlm}. 
	To this end, we adopt the Diffusion-LM architecture~\cite{li2022diffusionlm} and apply an embedding function $\bfe_{\phi}(\cdot)$, which maps each token onto a continuous vector in $\mathbb{R}^d$. Given a shell geometry, its corresponding implicit equation is tokenized as a sequence $\bfw = [w_1, w_2, \ldots, w_n]$. The $i$-th token $w_i$ may correspond to a function, operator, or constant. Each token then undergoes an embedding to yield the embedding of the sequence as 
	\begin{equation}
		\bfe_{\phi}(\bfw)=[\bfe_{\phi}(w_1), \bfe_{\phi}(w_2), \ldots, \bfe_{\phi}(w_n)]^T \in \mathbb{R}^{n \times d},
	\end{equation}
	where $d$ is a hyperparameter denoting the embedding dimension. 
	The reverse process aims to recover the original input embeddings through iterative denoising, which are subsequently mapped back to the discrete sequence $\bfw$ via a trainable rounding step (applying a softmax function at each position and selecting the most likely token). In this work, we jointly train the diffusion model and embedding function end-to-end, which allows the embedding space to evolve alongside the diffusion process to improve generation quality.
	
\begin{figure}[t!]
		\centering
		\includegraphics[width = \textwidth]{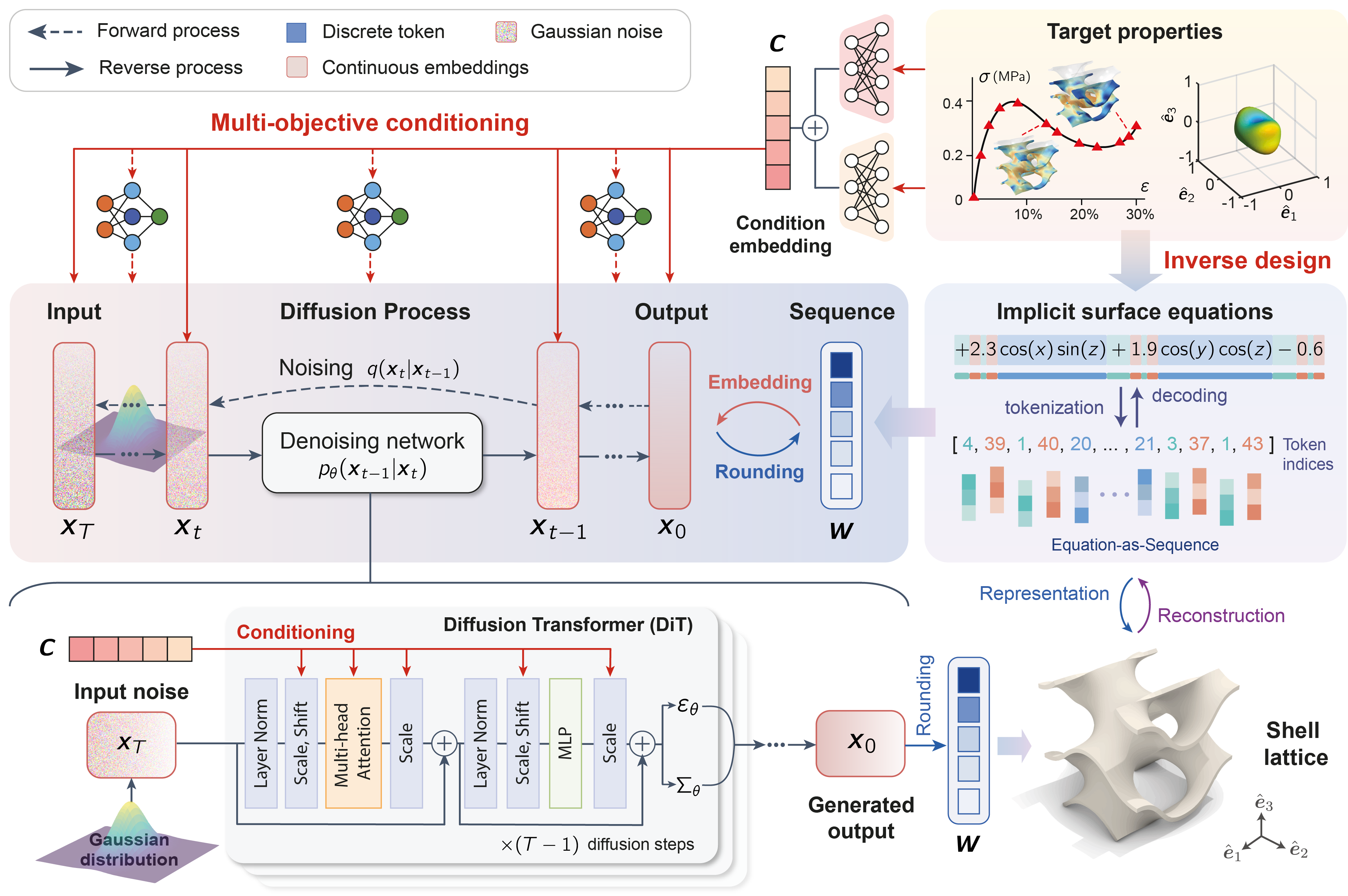}
		\captionsetup{justification=justified}
		\caption{\textbf{Schematic overview of the DiffuMeta framework for the inverse design of shell metamaterials with target mechanical properties.} 
		The process begins with discrete equation sequences representing implicit surface geometries, which are converted into a continuous space. During the forward (noising) process, Gaussian noise is progressively added to the embeddings over $T$ timesteps. The reverse (denoising) process employs a diffusion transformer to iteratively remove noise and reconstruct the original embeddings. To enable property-conditioned generation, target mechanical properties (stress-strain responses $\boldsymbol{\sigma}$ and homogenized stiffness tensor $\mathbb{C}$) are encoded into conditioning embeddings and integrated into the denoising network through cross-attention and adaptive layer normalization (AdaLN) mechanisms. Following denoising, a trainable rounding step maps the continuous outputs back to discrete token sequences, which are then decoded to generate the corresponding shell geometries with desired mechanical characteristics.}
		\label{fig:diffusion-model}
	\end{figure}

	As shown in Fig.~\ref{fig:diffusion-model}, our DiffuMeta framework formulates the inverse design of shell metamaterials as a conditional denoising diffusion process. The forward process progressively adds Gaussian noise to the embeddings over $T$ timesteps. During reverse denoising, the diffusion transformer iteratively removes this noise from randomly sampled noisy embeddings to reconstruct physically meaningful implicit equations. To guide the generation toward desired properties, we employ classifier-free guidance \cite{ho2022classifierfree}, which directly incorporates conditioning labels $\bfc$ without requiring auxiliary classifiers (Methods). As the backbone denoising network, we adopt a transformer architecture adapted from diffusion transformers~\cite{peebles2023scalablea} (DiT) to learn the reverse process. The transformer's ability to capture sequential dependencies is particularly well-suited for modeling implicit equations, where small perturbations can significantly impact the resulting shell geometry and its properties. The target properties, including the stress responses $\boldsymbol{\sigma}$ and the homogenized elastic properties (e.g., effective Poisson's ratios), are projected to high-dimensional embeddings by (learnable) linear layers, which are then integrated into the DiT model via cross-attention mechanisms~\cite{vaswani2017attention} and adaptive layer norm~\cite{peebles2023scalablea} (adaLN) blocks as conditioning labels (Supplementary Sections 3 and 4).
	
	\subsection*{Guided design of shell metamaterials}
	
	\begin{figure}[!htbp]
		\centering
		\includegraphics[width = 0.9\textwidth]{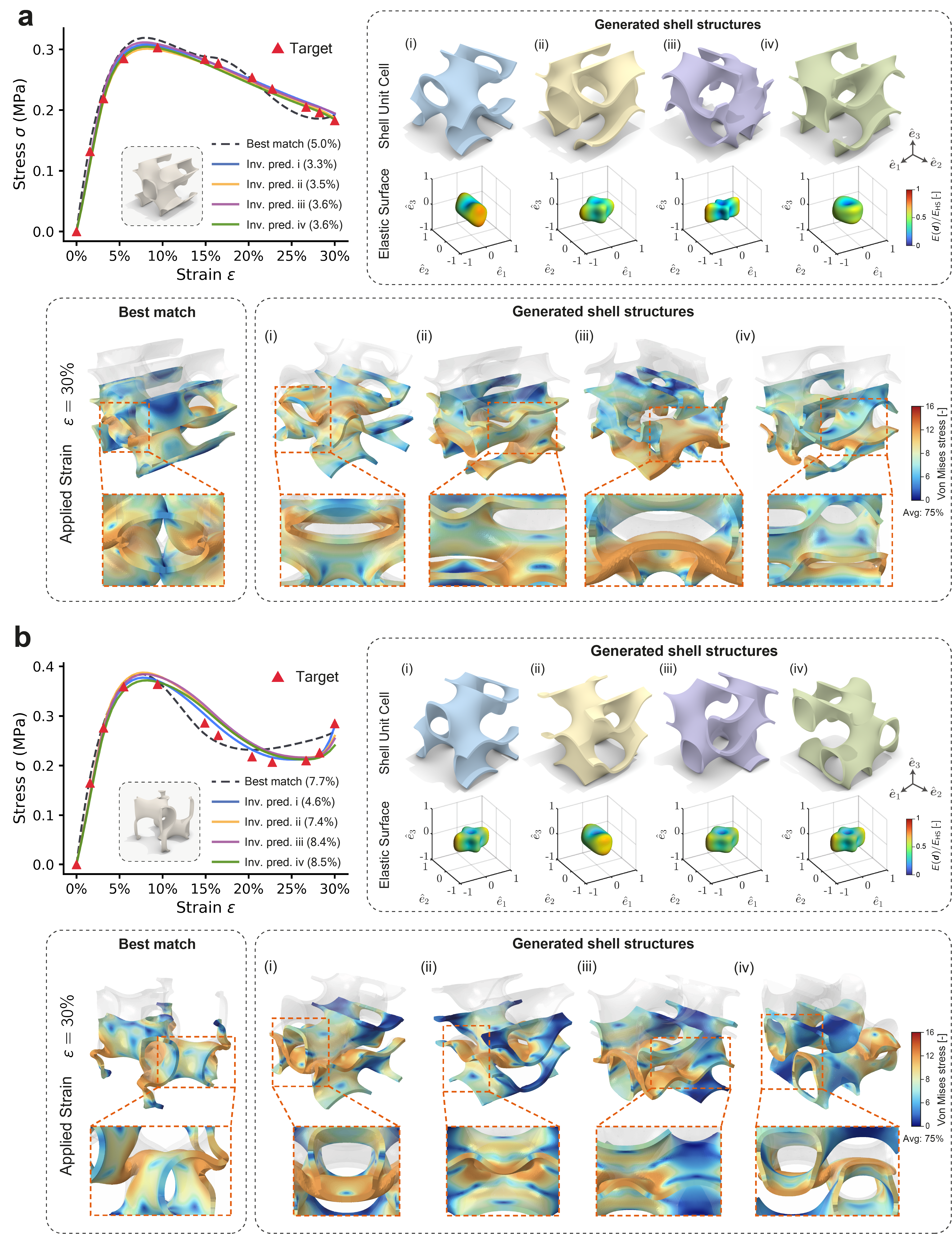}
		\captionsetup{justification=justified}
		\caption{\small \textbf{Inverse design of shell metamaterials with target compressive stress-strain responses.} The model is conditioned on target responses (denoted by red triangles) exhibiting \textbf{(a)}~pronounced softening and \textbf{(b)}~initial softening and subsequent hardening. Generated designs (\romannumeral 1)-(\romannumeral 4)~show diverse geometric configurations that achieve similar mechanical behavior through different deformation mechanisms. Each example shows the deformed unit cells of both the best match from the training dataset and the generated designs at $30\%$ strain, revealing varying stress distributions and structural responses. Elastic surface plots demonstrate distinct anisotropic 3D elastic stiffnesses for each generated design. All shown stress-strain responses and elastic surfaces were obtained via FE analysis, with NRMSE values (in brackets) quantifying design accuracy relative to target responses.}
		\label{fig:results-one-to-many}
	\end{figure}
	
	We first demonstrate that our DiffuMeta framework successfully captures the underlying data distribution and generates unseen, physically realizable shell structures. We generate $200$ implicit equations in an unconditional setting and evaluate the generation quality using three key metrics: (1) \textit{validity} -- the percentage of valid implicit equations producing single-connected surfaces without disjoint components; (2) \textit{novelty} -- the percentage of valid shell geometries not found in the training dataset; (3) \textit{uniqueness} -- the percentage of unique shell structures among the valid shells. {Our model achieves an overall validity of $74.0\%$, a novelty score of $100\%$, and a uniqueness score of $100\%$. Notably, this validity rate far exceeds the $3.2\%$ validity rate achieved by random sampling of implicit equations within the design space (Supplementary Section 6.2), highlighting the model's ability to avoid trivial solutions while effectively generating diverse, physically meaningful structures.} The generated designs exhibit diverse mechanical behaviors and distinct novel topologies (Supplementary Fig.~8), enabling efficient generation of diverse valid designs at negligible computational cost. We emphasize that the mapping from discrete implicit equations to continuous surface geometries is inherently challenging -- no existing method directly evaluates the quality of resulting surface geometries from symbolic equations alone. The high validity and uniqueness scores confirm the model's ability to capture the complex shell design space, while the high novelty rate underscores its ability to extrapolate beyond the training data.
	
	Next, we evaluate DiffuMeta's performance in the guided generation of shell designs that match target nonlinear stress-strain responses. As a first example (Fig.~\ref{fig:results-one-to-many}a), we select a target response with pronounced softening after reaching a peak stress. The designs proposed by DiffuMeta closely match the target response, achieving normalized root mean squared error (NRMSE; {$e$}) values of the FE-calculated response versus target response ranging from $3.3\%$ to $3.6\%$ (compared with $5.0\%$ for the closest existing design in the training dataset). Second, we consider a target response with notable softening at approximately $8\%$ applied strain, followed by hardening and a gradual stress increase as the structure compresses further (above $25\%$; Fig.~\ref{fig:results-one-to-many}b). While the best match in the training dataset achieves an NRMSE of $e = 7.7\%$, our inverse-designed solutions show improved performance, with the best achieving $4.6\%$ and variations ranging up to $8.5\%$. The generated shell structures successfully capture the characteristic stress increase at higher strains by leveraging contact interactions between structural elements, demonstrating the model's ability to predict complex deformation mechanisms under large deformations. {Compared with direct dataset search, our method generates multiple novel designs not present in the training data while achieving the same target properties, and more importantly, at negligible computational cost ($\sim$0.2 seconds per design on a single GPU; Supplementary Section 5).} Notably, the generated structures in Fig.~\ref{fig:results-one-to-many}a(\romannumeral 1) and Fig.~\ref{fig:results-one-to-many}b(\romannumeral 1) have nearly identical geometries, differing only in minor variations of local pore morphology and curvature, yet they produce substantially different mechanical responses. The corresponding implicit equations for generated shell designs are provided in Supplementary Section 6.3, revealing that their geometric similarity cannot be inferred from their drastically different mathematical formulations. This contrast underscores the critical challenge in the inverse design of nonlinear metamaterials, where subtle geometric variations can lead to significant changes in mechanical behavior due to the complex structural interactions -- while DiffuMeta efficiently captures this intricate structure-to-property mapping.
	
	A further key advantage of DiffuMeta lies in its ability to generate multiple distinct designs for the same target response, addressing the fundamental one-to-many mapping challenge in inverse design problems (i.e., multiple designs can produce similar mechanical behavior). The overall performance distributions show that the model consistently generates multiple designs whose properties closely match the target responses (Supplementary Fig.~6), demonstrating stable and robust inverse design performance. Unlike traditional optimization methods such as topology optimization, which typically converge to a single solution~\cite{papadopoulos2021computing}, our generative approach naturally produces diverse candidates that all satisfy the specified targets. Notably, the generated structures exhibit diverse geometries (e.g., variations in surface curvature, pore distribution patterns, and local feature dimensions), while they consistently reproduce key mechanical properties, such as initial stiffness, softening, plateau stress, and hardening, through distinct deformation mechanisms, as evidenced by the von Mises stress distribution plots. In addition, the elastic surface plots of generated designs in Fig.~\ref{fig:results-one-to-many}a-b reveal that each design exhibits unique anisotropic elastic properties, offering potential design flexibility for applications requiring secondary performance criteria. 
	
	\subsection*{Inverse design of shell metamaterials for unseen target properties}
	
	\begin{figure}[!t]
		\centering
		\includegraphics[width = \textwidth]{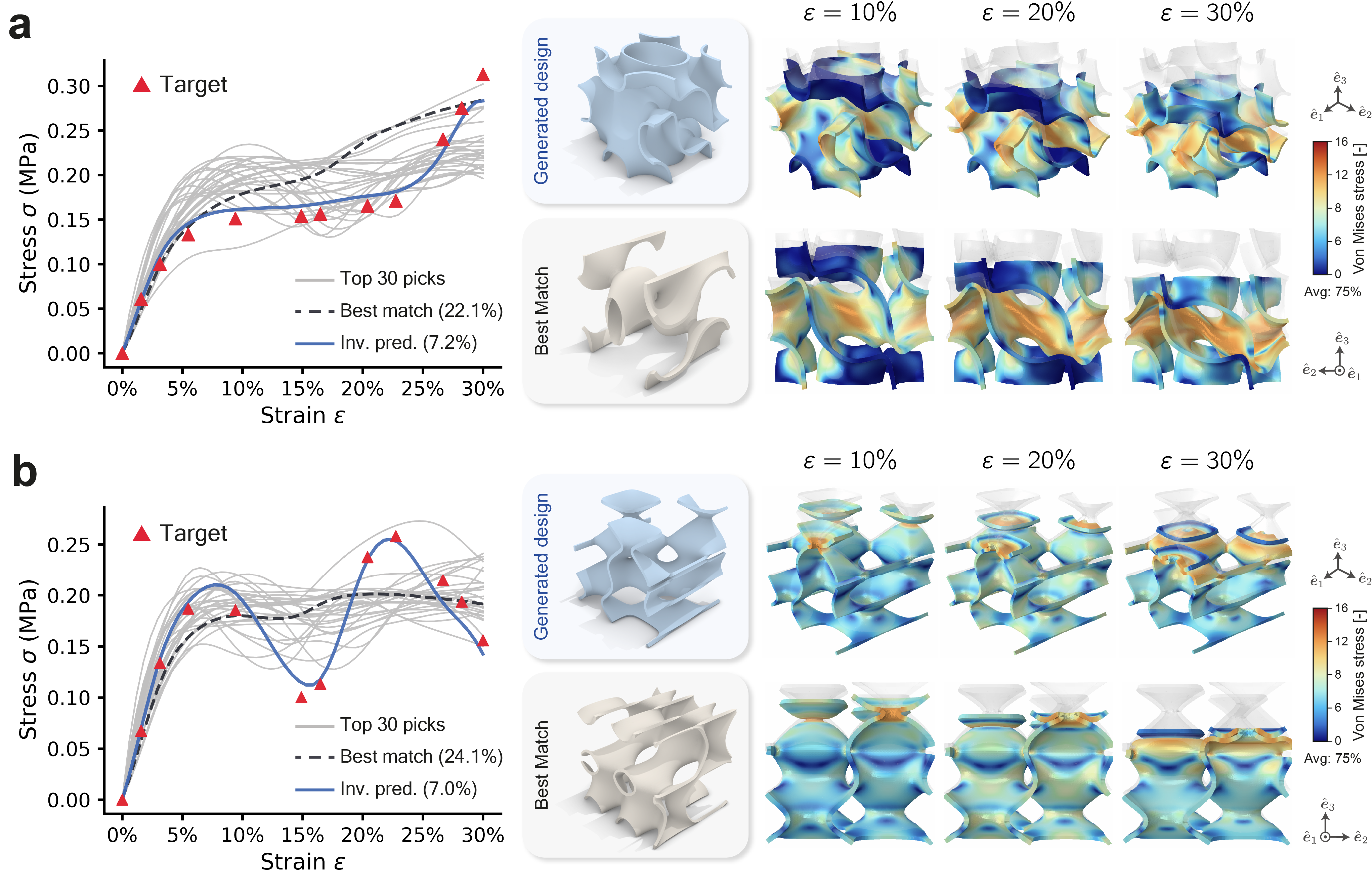}
		\captionsetup{justification=justified}
		\caption{ \textbf{Inverse-designed shell metamaterials with unseen stress-strain responses}. The diffusion model generates designs conditioned on target stress-strain curves (denoted by red triangles) exhibiting \textbf{(a)}~a stress plateau and subsequent hardening; \textbf{(b)}~pronounced buckling-induced stress peaks at 8\% and 22\% strain. These responses significantly extend beyond the training distribution. We highlight the top 30 closest matches from the training set in gray in \textbf{(a)} and \textbf{(b)}, selected based on normalized root mean square error (NRMSE), which deviate substantially from the target. All shown stress-strain responses and elastic surfaces were obtained via FE analysis, with NRMSE values (in brackets) quantifying design accuracy relative to target responses.}
		\label{fig:results-extrapolation}
	\end{figure}
	
	So far, we have demonstrated DiffuMeta's capability to generate novel shell designs conditioned on complex stress-strain responses. While the target responses in Fig.~\ref{fig:results-one-to-many} lie within the training distribution and can be reasonably matched by existing designs, we next assess the model's extrapolation ability by examining two target stress-strain curves that exhibit highly nonlinear, non-monotonic compressive responses, significantly deviating from the training distribution. The target curve in Fig.~\ref{fig:results-extrapolation}a shows a stress plateau followed by strain hardening. The second target response (Fig.~\ref{fig:results-extrapolation}b) exhibits two pronounced stress peaks. In both cases, the best matches from the training dataset poorly reconstruct the target response ($e=22.1\%$ and $e=24.1\%$, respectively), which is unsurprising -- indicating the vastness of the design space (and the small training set). We also show the top 30 candidate designs from the training dataset, which generally deviate from the target, particularly at large strains. In contrast, the model identifies designs that accurately reproduce the complex target responses (achieving $e=7.2\%$ and $e=7.0\%$, respectively). 
	
	As shown in Fig.~\ref{fig:results-extrapolation}a, the generated structure initially undergoes progressive shell bending, followed by contact between opposing surfaces, which creates new internal load paths and stiffens the structure, producing the characteristic stress rise at large strains. In contrast, the design in Fig.~\ref{fig:results-extrapolation}b exhibits a sequential deformation mechanism characterized by initial elastic buckling of its curved surfaces ($\varepsilon < 15\%$), which causes softening, followed by the development of multiple contact interfaces, which leads to hardening, and finally another local buckling instability that results in the second softening phase at around $22\%$ applied strain. Remarkably, these highly nonlinear stress-strain responses of shell lattices at finite strains emerge from intricate structural interactions rather than being explicitly encoded in their implicit equations. The mapping from abstract mathematical representations, which define only the initial topology, to these sophisticated nonlinear responses is therefore challenging. Nevertheless, DiffuMeta successfully learns this intricate relationship by identifying mathematical patterns that correlate geometric features with the resulting deformation characteristics of various shell metamaterials.
	
	\subsection*{Multi-target conditional generation of shell metamaterials}
	
	\begin{figure}[!htbp]
		\centering
		\includegraphics[width = 0.7\textwidth]{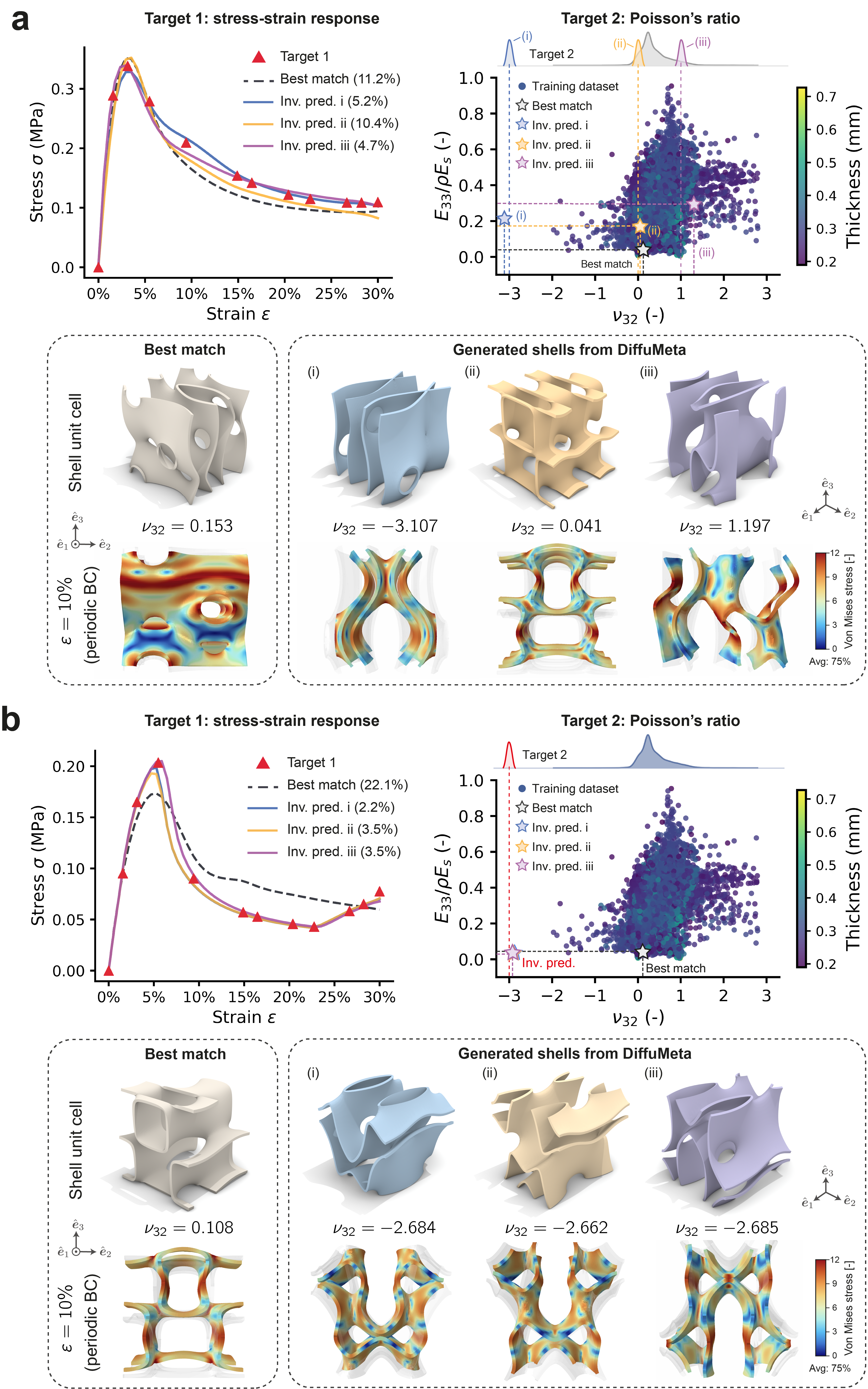}
		\caption{\small\textbf{Multi-target conditional generation of shell metamaterials.} \textbf{(a)}~Generated shell designs conditioned on a target stress-strain response while achieving three distinct target effective (anisotropic) Poisson's ratios {(measured at small strains)}: $\nu_{32} = -3.0$, $\nu_{32} = 0.0$, and $\nu_{32} = 1.0$. \textbf{(b)}~Inverse design for unseen target property combinations, where both the stress-strain response and Poisson's ratio ($\nu_{32} = -3.0$) extend well beyond the training distribution. 
		Each example shows the achieved Poisson's ratio of the best match from the training dataset and the generated designs, compared to the distribution of training samples in the relevant property space (each dot represents a unit cell in the training data, color-coded by shell thickness). The effective directional Young's modulus $E_{33}$ is normalized by the Young's modulus of the base material $E_s$ and the relative density $\rho$. The simulated samples at 10\% compressive strain illustrate their lateral deformations. All shown stress-strain responses were obtained via FE analysis, with NRMSE values (in brackets) quantifying design accuracy relative to target responses.
		}
		\captionsetup{justification=justified}
		\label{fig:results-multi-objective}
	\end{figure}
	
	To demonstrate our framework's ability to extend beyond a single design objective, we condition the generation of shell metamaterials on multiple target properties simultaneously (Methods). We focus on the task of jointly tuning both linear and nonlinear mechanical responses, e.g., the effective (anisotropic) Poisson's ratio and the nonlinear compressive stress-strain response. Such design targets can be leveraged for applications in, e.g., protective gear or biomedical implants, where a specific nonlinear stress-strain response (such as controlled buckling or energy absorption) is required, while also ensuring, e.g., auxetic behavior under small-strain compression to enhance conformability and mechanical stability. 
	
	As a first example, we consider a target stress-strain response exhibiting significant softening under compression, while simultaneously targeting three distinct effective Poisson's ratios {(measured at small strains)}: $\nu_{32} = -3.0$ (strongly auxetic), $\nu_{32} = 0.0$ (zero transverse strain), and $\nu_{32} = 1.0$ (high positive Poisson's ratio). As shown in Fig.~\ref{fig:results-multi-objective}a, the generated shell structures not only closely reproduce the target stress-strain responses (ranging from $e=4.7\%$ to $e=10.4\%$) but also exhibit the intended Poisson effect. Fig.~\ref{fig:results-multi-objective}a further compares the homogenized response of the closest design in the training set ($e=11.2\%$) with their deformed states at 10\% compressive strain under periodic boundary conditions. In contrast to the best match, which exhibits moderate lateral expansion, the generated designs demonstrate remarkably diverse lateral deformation behaviors. The auxetic design Fig.~\ref{fig:results-multi-objective}a($\romannumeral 1$) undergoes significant lateral contraction upon compression, achieving the target negative Poisson's ratio through inward bending of its curved surfaces. Design Fig.~\ref{fig:results-multi-objective}a($\romannumeral 3$) shows pronounced lateral expansion, while design Fig.~\ref{fig:results-multi-objective}a($\romannumeral 2$) maintains nearly constant lateral dimensions. The corresponding von Mises stress distribution plots further reveal that each geometry accommodates the applied compressive load through distinct deformation pathways and stress localization patterns to maintain the same target stress-strain responses while achieving various Poisson's ratios targets. (It is remarkable that the linear Poisson's ratio, a measure valid at small strains, here serves well as an effective descriptor also for the nonlinear, large-strain lateral sample deformation.)
	
	As a second example, we consider a target stress-strain response with significant softening followed by stiffness increase at higher strains (Fig.~\ref{fig:results-multi-objective}b), which lies outside the training distribution, as evidenced by the poor match ($e = 22.1\%$) with the best training dataset candidate. Despite this challenging unseen target, the inverse-designed solutions closely follow the target curve and achieve significantly lower errors ($2.2\%$–$3.5\%$). In addition, the inverse-designed solutions successfully produce the desired extreme auxetic behavior, exhibiting pronounced lateral contraction under compression. Notably, the target Poisson's ratio of $\nu_{32} = -3.0$ also lies far outside the training distribution, as evidenced by the scatter plot of effective modulus versus Poisson's ratio for all training samples. {Notably, the generated designs achieving exceptional properties (Fig.~\ref{fig:results-extrapolation}-\ref{fig:results-multi-objective}) remain strictly within the defined design space bounds, with all coefficient values within the training range (Supplementary Section 6.3). This demonstrates that DiffuMeta achieves \emph{extrapolation in the property space} through \emph{interpolation in the design space} by capturing the complex underlying structure-property relationships governing highly nonlinear behaviors, including plasticity, buckling, contact, etc.} Collectively, these results demonstrate the robust generalization capability of the proposed framework, enabling the inverse design of shell metamaterials with highly nontrivial combinations of unseen linear and nonlinear mechanical properties, surpassing the limits of the training distribution.
	
	\subsection*{Experimental validation}
	\begin{figure}[!htbp]
		\centering
		\includegraphics[width = \textwidth]{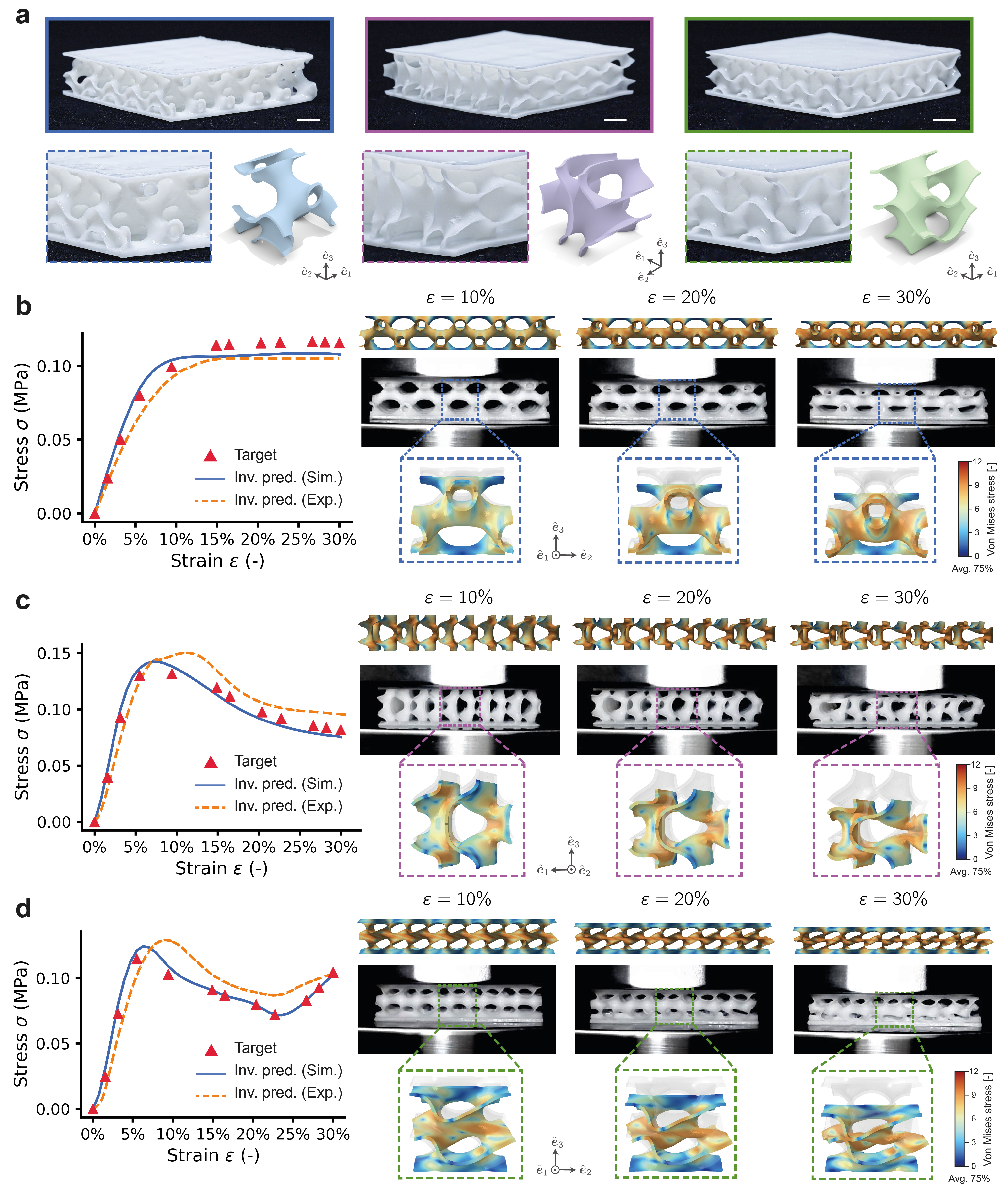}
		\caption{\small\textbf{Experimental validation of shell metamaterials generated by DiffuMeta.} \textbf{(a)}~Fabricated representative shell samples alongside their corresponding Computer-Aided Design (CAD) models. Samples are arranged in $5 \times 5 \times 1$ arrays with $10\%$ relative density. All scale bars are \SI{10}{\milli\meter}. \textbf{(b-d)}~Comparison between experimental stress-strain responses (dashed lines) and finite element simulation results (solid lines) for three distinct shell designs generated by DiffuMeta. 
		Target responses used as conditions for the diffusion model are indicated with red triangles. The deformed configurations of generated shell unit cells obtained from FE analysis at different strain levels are shown in $5 \times 5 \times 1$ tessellations for visualization. The experimental results demonstrate overall good agreement with both the target responses and simulation predictions for the nonlinear behavior, validating the effectiveness of the inverse design framework. 
		}
		\captionsetup{justification=justified}
		\label{fig:results-experiment}
	\end{figure}
	
	To validate our inverse design framework, we conduct experimental compression tests on 3D-printed shell samples generated by DiffuMeta (Methods). We select several distinct target responses from the test dataset (unseen during training) and fabricate the corresponding generated designs (Fig.~\ref{fig:results-experiment}a). Fig.~\ref{fig:results-experiment}b-d presents uniaxial compression test data in comparison with FE results. The experimental validation demonstrates overall good agreement between simulated and measured stress-strain responses, validating both the FE modeling accuracy and the robustness of our inverse design framework. The design in Fig.~\ref{fig:results-experiment}b shows an extended stress plateau up to high strains, maintaining structural integrity through its smooth, continuous surfaces without significant buckling instabilities. In contrast, Fig.~\ref{fig:results-experiment}c demonstrates pronounced softening, characterized by localized surface buckling. Fig.~\ref{fig:results-experiment}d illustrates a more complex target response featuring initial softening followed by hardening, which results from progressive contact formation between internal surfaces under large deformations. Notably, the fabricated samples successfully reproduce the complex nonlinear mechanical behaviors predicted by DiffuMeta, including a stable plateau, buckling-induced softening, and hardening due to contact. While some discrepancies exist between simulation and experiment due to, e.g., material variations, geometric imperfections, and fabrication-induced residual stresses, the overall stress-strain responses and key mechanical features are well captured experimentally. This close agreement validates the robustness and practical applicability of our inverse design framework, opening new possibilities for creating novel materials with tailored nonlinear responses for applications in soft robotics, morphing structures, and energy-absorbing systems.
	
	\section*{Discussion}
	The presented generative modeling framework enables the inverse design of architected materials with precisely targeted combinations of nonlinear stress-strain responses and linear elastic properties, demonstrating robust extrapolation beyond the training domain. DiffuMeta employs a novel algebraic language-based parameterization that represents implicit equations as discrete token sequences. This provides a flexible, low-dimensional representation encoding a vast shell design space, while preserving intricate topological features essential for their mechanical functionality. By integrating these sequences into a diffusion-based framework, DiffuMeta effectively learns how subtle variations in implicit equations lead to significant changes in mechanical performance, capturing the complex mappings between abstract implicit equations and highly non-trivial deformation mechanics, including buckling and contact.
	
	A key advantage of our approach lies in the probabilistic nature of the diffusion model, which inherently addresses the ill-posedness of the inverse design problem by generating multiple solutions for a given target response, unlike conventional optimization methods that converge to a single solution and depend heavily on initialization. In addition, the framework demonstrates strong generalization, successfully identifying designs for highly non-trivial unseen target responses. Furthermore, our approach enables multi-objective conditional generation, simultaneously controlling linear and nonlinear properties while effectively addressing the complex coupling between different mechanical responses. 
	
	We note that our framework requires target properties to be physically achievable within the shell metamaterial design space. {While the model successfully generates designs that extend beyond existing property boundaries (Figs.~\ref{fig:results-extrapolation}-\ref{fig:results-multi-objective}), this extrapolation capability is inherently bounded by the physics encoded in the training data; targets requiring mechanical mechanisms not represented in the dataset would necessitate expanding the training data to include richer mechanical responses (e.g., hierarchical designs, multi-stage plasticity, or snap-through instabilities).} Furthermore, the current approach does not explicitly incorporate physical constraints during generation, which may occasionally yield designs that, while mathematically valid, are challenging to fabricate. Future work could integrate physics-informed constraints directly into the diffusion process~\cite{bastek2025physicsinformed} to enhance robustness and physical validity. Additionally, active learning strategies~\cite{settles2009active} could be employed to iteratively sample underrepresented regions of the design space, improving generation quality for out-of-distribution targets. The versatility and efficiency of DiffuMeta make it readily available for extension to other metamaterial architectures~\cite{zheng2023unifying} and {targeting alternative properties, such as impact absorption~\cite{deng2022inverse,perroni-scharf2025dataefficienta}, optical responses~\cite{li2025inversea}, and multiphysics properties~\cite{mirabolghasemi2019thermal},} by altering the training dataset accordingly. The framework could also be integrated with large language models to enable natural language-driven design specifications~\cite{park2025multimodal}, further enhancing the design of multifunctional metamaterials. These extensions underscore the potential of our generative modeling approach, offering a scalable and generalizable framework for the inverse design of diverse metamaterial systems with unprecedented capabilities.
	
	
	\section*{Methods}
	
	We here provide details of the diffusion model derivation and architecture, the FE simulation setup to evaluate the stress-strain response of shell lattices, and fabrication and experimental characterization details. Additional explanations can be found in the Supplementary Information.
	
	\label{sec:methods}
	\subsection*{Denoising diffusion models}
	At its core, a diffusion model consists of two fundamental processes: forward and reverse. The forward process systematically perturbs the input data $\bfx_0 \sim q(\bfx)$ by gradually adding Gaussian noise over $T$ steps, transforming it into a standard Gaussian noise $\bfx_T \sim \mathcal{N}(\bfnull,\bfI)$. This noise injection follows a fixed Markov chain, in which each transition from $\bfx_{t-1}$ to $\bfx_t$ is governed by
	\begin{equation}\label{eq:forward-process}
		q(\bfx_{t}| \bfx_{t-1}) = \mathcal{N}(\bfx_t; \sqrt{1-\beta_t}\bfx_{t-1}, \beta_t \bfI),
	\end{equation}
	where the hyperparameter $\{\beta_t \in (0,1)\}_{t=1}^T$ regulates the amount of noise added at diffusion step $t$. We can then sample $\bfx_t$ at any time step $t$ during the forward process by 
	\begin{equation}\label{eq:q-sample}
		\begin{aligned}
			q(\bfx_{t}| \bfx_0) &= \mathcal{N}(\bfx_t; \sqrt{\Bar{\alpha}_t}\bfx_0, (1-\Bar{\alpha}_t)\bfI), \quad, \text{where} \quad \alpha_t = 1 - \beta_t, \Bar{\alpha}_t = \prod_{i=1}^t \alpha_i\\
			\bfx_t &= \sqrt{\Bar{\alpha}_t}\bfx_0 + \sqrt{1-\Bar{\alpha}_t}\varepsilon, \quad \varepsilon \sim \mathcal{N}(\bfnull, \bfI).
		\end{aligned}
	\end{equation}
	The reverse denoising process seeks to reconstruct the original data $\bfx_0$ by first sampling from $q(\bfx_T)$ and then proceeding in reverse time direction, sampling from $q(\bfx_{t-1}|\bfx_t)$ until $\bfx_0$. Since the true denoising transition $q(\bfx_{t-1}|\bfx_t, \bfx_0)$ is intractable, we can approximate it using a neural network $p_{\theta}(\bfx_{t-1}| \bfx_{t})$ parameterized by $\theta$. The reverse process $p_{\theta}(\bfx_{0:T})$ is characterized as a Markov chain as
	\begin{equation}\label{eq:reverse-process}
	\begin{aligned}
		p_{\theta}(\bfx_{0:T})&=p(\bfx_T)\prod^T_{t=1}p_{\theta}(\bfx_{t-1}|\bfx_t),\quad p_{\theta}(\bfx_{t-1}| \bfx_{t}) = \mathcal{N}(\bfx_{t-1};\mu_{\theta}(\bfx_t, t), \sigma_t^2 \bfI),
	\end{aligned}
	\end{equation}
	where $p(\bfx_T)=\mathcal{N}(\bfx_T; \bfnull, \bfI)$ denotes the standard Gaussian distribution, $\mu_{\theta}(\cdot)$ is the predicted mean of the Gaussian transition $q(\bfx_{t-1}| \bfx_{t})$, and $\sigma_t^2$ is a variance depending on the time step: $\sigma_t^2=\dfrac{1-\Bar{\alpha}_{t-1}}{1-\Bar{\alpha}_t}\beta_t$. $\bfx_{t-1}$ can be sampled as
	\begin{equation}
	\bfx_{t-1} = \mu_{\theta}(\bfx_t,t)+\sigma_t \boldsymbol{\varepsilon}, \quad \text{with}\quad \boldsymbol{\varepsilon}\sim\mathcal{N}(\bfnull, \bfI).
	\end{equation}
	To train the diffusion model, the objective is to maximize the (log-)likelihood $\log p_{\theta}(\bfx_0)=\int p_{\theta}(\bfx_{0:T})\mathrm{d}\bfx_{1:T}$, which is computationally infeasible. Instead, we maximize the tractable variational lower bound of the log-likelihood. The loss funtion $\mathcal{L}$ can be rewritten as 
	\begin{equation}
		\mathcal{L}(\theta) = \mathbb{E}\left[ \underbrace{\mathcal{D}_{\text{KL}}(q(\bfx_T|\bfx_0) \| p(\bfx_T))}_{\mathcal{L}_T} + \sum^T_{t=2} \underbrace{\mathcal{D}_{\text{KL}}(q(\bfx_{t-1}|\bfx_t,\bfx_0 )\| p_{\theta}(\bfx_{t-1}|\bfx_t))}_{\mathcal{L}_{t-1}} \underbrace{- \log p_{\theta}(\bfx_0|\bfx_1)}_{\mathcal{L}_0}
		\right],
	\end{equation}
	where $\mathcal{D}_{\text{KL}}(q \| p)$ denotes the Kullback-Leibler (KL) divergence between two distributions $q$ and $p$. 
	\subsection*{Diffusion transformer architecture}\label{subsec:methods-diffusion-models}
	To extend the continuous diffusion models to the text domain, we adopt the Diffusion-LM~\cite{li2022diffusionlm} framework that maps discrete text into a continuous latent space via an embedding function $\bfe_{\phi}$. The framework is then trained to jointly learns the diffusion model's parameters and word embeddings in an end-to-end manner. A detailed derivation of the end-to-end training objective is provided in Supplementary Section 3. To condition the model on a target, we adopt the classifier-free guidance approach~\cite{ho2022classifierfree}. The model is trained to jointly learn both conditional and unconditional generation by randomly replacing the condition label $\bfc$ with a learnable ``null'' token $\varnothing$ during training, with a fixed probability $p_{\text{drop}}$. This approach enables flexible control during inference by replacing the predicted noise $\boldsymbol{\varepsilon}_{\theta}$ with a weighted combination of conditional and unconditional predictions as
	\begin{equation} 
		\hat{\boldsymbol{\varepsilon}}_{\theta}(\bfx_t, t, \bfc) =  \omega \boldsymbol{\varepsilon}_{\theta}(\bfx_t, t, \bfc) + (1-\omega) \boldsymbol{\varepsilon}_{\theta}(\bfx_t, t, \varnothing), 
	\end{equation}
	where $\omega\geq1$ is the guidance weight that balances the influence of conditional scores, controlling the trade-off between generation performance and sampling diversity. Here, $\boldsymbol{\varepsilon}_{\theta}(\bfx_t, t, \bfc)$ and $\boldsymbol{\varepsilon}_{\theta}(\bfx_t, t, \varnothing)$ denote the predicted noise from the conditional and unconditional models, respectively. Note that our diffusion model is trained to directly predict $\bfx_0$ instead of noise. We can derive the noise predictions from the model output $f_{\theta}(\bfx_t, t)$ using Eq.~\ref{eq:q-sample} as $\boldsymbol{\varepsilon}_{\theta} = (\bfx_t - \sqrt{\Bar{\alpha}_t}f_{\theta}(\bfx_t, t))/\sqrt{1-\Bar{\alpha}_t}$.
	
	To learn the reverse process, we adopt a transformer architecture based on diffusion transformers~\cite{peebles2023scalablea} (DiTs). Unlike traditional diffusion models that use convolutional networks, a DiT leverages self-attention to capture long-range dependencies and complex interactions, making it effective for high-dimensional, structured data. To incorporate conditional labels $\bfc\in \mathbb{R}^h$ into our model, we first project each value into the embedding space, using a (learnable) linear projection layer and yielding the condition embeddings $\bfy \in \mathbb{R}^{d}$. When multiple conditional properties are considered, each is processed by a separate projection layer, and the resulting conditional embeddings are concatenated. The conditional information, processed through the projection layer, is then integrated into the denoising process of the transformer model via cross-attention mechanisms~\cite{vaswani2017attention}, enabling the model to effectively capture interactions between input and conditional information. Additionally, the conditional embeddings $\bfy$ are projected and added to the time embedding $\bft$, which is then processed through an adaptive layer norm~\cite{peebles2023scalablea} (adaLN) to predict a series of parameters, including global scale and shift parameters $\gamma$ and $\beta$, and dimension-wise scaling parameters $\alpha$. The adaLN blocks adaptively modulate the feature activations, leveraging the conditional information, which has been shown to be effective in improving the model’s generation performance~\cite{peebles2023scalablea}. {Besides the compressive response and linear elastic properties investigated in this paper, the framework can be readily extended to a broader range of target properties without architectural changes.} We refer to the code (see `Code availability') for technical details regarding our DiT architecture.
	
	\subsection*{Machine learning protocols}
	{
	The performance of the proposed generative framework depends on several key design and training choices. The dataset of $23{,}534$ samples balances the computational cost of FE simulations with model performance; ablation studies on smaller subsets indicate diminishing gains with additional data, while the full dataset yields robust inverse design performance. The transformer architecture (including the number of attention heads, embedding dimension, and transformer hidden dimension) and training hyperparameters (including optimizer, learning rate, and noise schedule) were determined through systematic ablation studies, which reveal that model performance is robust across a range of hyperparameter choices. Detailed ablation experiments and the final model configuration are provided in Supplementary Section 4. To comprehensively assess the model's generation performance, we evaluate the distribution of generated design properties under both conditional and unconditional generation settings (Supplementary Section 6). The model demonstrates stable and reliable generation performance across various target properties, effectively leveraging the probabilistic nature of generative models to address the one-to-many mapping inherent to inverse design.
	}
	
	\subsection*{Finite element simulation}
	
	To extract the stress-strain response of constructed shell structures, FE simulations are conducted using the commercial software ABAQUS/CAE 2023. The base material of all structures is described by an elastic-plastic constitutive model, with parameters calibrated from experimental uniaxial tensile test data (see ref.~\cite{rosa2025enhanced} for material characterization details). Each structure is discretized using a finite-strain, three-node triangular shell element (S3R). A periodic mesh is generated to ensure compatibility with the imposed periodic boundary conditions. Mesh refinement studies yield sufficient convergence of the effective (homogenized) stress-strain response and the stress distribution at an element size of $0.2 \mathrm{mm}$, as presented in Supplementary Section 2. The shell unit cell is positioned between two rigid plates with periodic boundary conditions prescribed on all lateral surfaces. The bottom plate is fixed, and a compressive strain up to $30\%$ is applied by displacing the top plate vertically down. General contact with a friction coefficient of 0.6 is applied to all interacting surfaces within the unit cell. To promote convergence and numerical stability throughout the large-deformation regime in the presence of buckling and contact, a dynamic implicit solver with moderate dissipation is employed to extract the quasistatic response. We record displacements, stresses, and reaction forces of nodes in contact with the upper rigid plate at 50 equidistant strain increments. To facilitate the ML model training, we reduce the dimensionality of the target by representing the stress-strain response as a vector $\boldsymbol{\sigma}\in\mathbb{R}^{11}$, which contains the compressive stress at 11 strategically selected strain levels across the applied strain range. We verify that the full stress-strain response can be accurately reconstructed via cubic spline interpolation of these sampled points (see Supplementary Section 2 for further details). To validate quasistatics, we ensure that the kinetic energy remains below $1\%$ of the internal energy at all strain increments. {The effective stiffness tensor is computed via FE homogenization of a single unit cell with periodic boundary conditions, and the effective (anisotropic) Poisson's ratios are extracted at small strains.} All simulations are performed on the Euler cluster of ETH Zurich.
	
	\subsection*{Fabrication and experimental validation}
	
	Specimens tested in this work were fabricated using digital light synthesis (DLS) with a Carbon M2 3D printer, operating at an average writing speed of \SI{22}{\milli\meter\per\hour} and a layer thickness of \SI{100}{\micro\meter}. The printing material is UMA 90 resin (Carbon, Inc.), a one-part photopolymer with a tensile modulus of \SI{484}{\mega\pascal}, ultimate strength of \SI{11.9}{\mega\pascal}, and 19.3\% elongation at break~\cite{rosa2025enhanced}. After printing, samples were first cleaned by immersion in isopropyl alcohol (IPA) under orbital agitation for 5~minutes to remove excess resin, followed by UV post-curing for 3~minutes to ensure complete photopolymerization. To enable direct comparison across different shell topologies, the wall thickness of each design was scaled proportionally to maintain a consistent relative density of 10\% for all specimens. Two rigid plates of thickness \SI{1}{\milli\meter} are added at the top and bottom during 3D printing to facilitate a uniform load distribution during compression testing. Uniaxial displacement-controlled compression tests were conducted using an Instron 5943 Single Column testing machine at a quasistatic strain rate of \SI{0.0005}{\per\second}. The effective strains and stresses were calculated from the recorded force-displacement data and measured sample dimensions. Load-displacement data were acquired at a sampling frequency of \SI{10}{\hertz} throughout the compression process.
	
	\section*{Data availability}
	The training and test dataset (consisting of implicit equations, their corresponding stress-strain responses, and effective stiffness data) and the pre-trained models are available in the ETHZ Research Collection{~\cite{DiffuMeta2025Dataset}}.
	
	\section*{Code availability}
	{The complete source code for this work, including data generation, model training, and the conditional generation of new metamaterial designs with target responses, is available at} \url{https://github.com/li-zhengz/DiffusionMetamaterials.git} and on Zenodo~\cite{DiffuMeta2026Code}.
	
	\section*{Acknowledgments}
	We thank Dr. Prakash Thakolkaran for his early contributions that informed the development of the design space and the parameterization. This research received financial support from ETH Zurich through the ETH+ grant SynMatLab.
	
	\section*{Authors contributions}
	\textbf{Li Zheng:} \\
	\textbf{Siddhant Kumar:} \\
	\textbf{Dennis M. Kochmann:}
	
	\section*{Competing interests} 
	The authors declare no competing interests.
	
	\section*{Figure captions}
	\vspace{2mm}
	\textbf{Figure 1}: \textbf{Overview of the shell-based metamaterial parameterization and design space generation process.} \textbf{(a)}~A shell lattice is generated from the implicit level set equation, which can be tokenized into a sequence of discrete mathematical tokens drawn from a structured vocabulary. Novel shell designs can be generated by sampling and recombining these tokens. \textbf{(b)}~To obtain the stress-strain responses, we conduct finite element (FE) simulations by imposing rigid plates on the top and the bottom with periodic boundary conditions on the lateral surfaces. A quasi-static compressive strain of up to 30\% along the $\hat\bfe_3$-direction is applied. The overall effective stress-strain response is extracted from the reaction forces. A representative deformed shell unit cell is shown at an applied strain of $30\%$. Also shown is the 3D elastic surface plot illustrating the directional effective Young's modulus (obtained via FE homogenization). \textbf{(c)}~Stress-strain responses of 200 structures randomly drawn from the dataset, and selected representative examples with distinct mechanical behaviors. Also shown are the photos of fabricated shell structures, each obtained by arranging corresponding shell unit cells in a $5 \times 5 \times 1$ array, with a total size of $50\mathrm{mm} \times 50\mathrm{mm} \times 10\mathrm{mm}$. All scale bars indicate $10\mathrm{mm}$. 
	
	\textbf{Figure 2}: \textbf{Schematic overview of the DiffuMeta framework for the inverse design of shell metamaterials with target mechanical properties.} The process begins with discrete equation sequences representing implicit surface geometries, which are converted into a continuous space. During the forward (noising) process, Gaussian noise is progressively added to the embeddings over $T$ timesteps. The reverse (denoising) process employs a diffusion transformer to iteratively remove noise and reconstruct the original embeddings. To enable property-conditioned generation, target mechanical properties (stress-strain responses $\boldsymbol{\sigma}$ and homogenized stiffness tensor $\mathbb{C}$) are encoded into conditioning embeddings and integrated into the denoising network through cross-attention and adaptive layer normalization (AdaLN) mechanisms. Following denoising, a trainable rounding step maps the continuous outputs back to discrete token sequences, which are then decoded to generate the corresponding shell geometries with desired mechanical characteristics.
	
	\textbf{Figure 3}: \textbf{Inverse design of shell metamaterials with target compressive stress-strain responses.} The model is conditioned on target responses (denoted by red triangles) exhibiting \textbf{(a)}~pronounced softening and \textbf{(b)}~initial softening and subsequent hardening. Generated designs (\romannumeral 1)-(\romannumeral 4)~show diverse geometric configurations that achieve similar mechanical behavior through different deformation mechanisms. Each example shows the deformed unit cells of both the best match from the training dataset and the generated designs at $30\%$ strain, revealing varying stress distributions and structural responses. Elastic surface plots demonstrate distinct anisotropic 3D elastic stiffnesses for each generated design. All shown stress-strain responses and elastic surfaces were obtained via FE analysis, with NRMSE values (in brackets) quantifying design accuracy relative to target responses.
	
	\textbf{Figure 4}: \textbf{Inverse-designed shell metamaterials with unseen stress-strain responses}. The diffusion model generates designs conditioned on target stress-strain curves (denoted by red triangles) exhibiting \textbf{(a)}~a stress plateau and subsequent hardening; \textbf{(b)}~pronounced buckling-induced stress peaks at 8\% and 22\% strain. These responses significantly extend beyond the training distribution. We highlight the top 30 closest matches from the training set in gray in \textbf{(a)} and \textbf{(b)}, selected based on normalized root mean square error (NRMSE), which deviate substantially from the target. All shown stress-strain responses and elastic surfaces were obtained via FE analysis, with NRMSE values (in brackets) quantifying design accuracy relative to target responses.
	
	\textbf{Figure 5}: \textbf{Multi-target conditional generation of shell metamaterials.} \textbf{(a)}~Generated shell designs conditioned on a target stress-strain response while achieving three distinct target effective (anisotropic) Poisson's ratios {(measured at small strains)}: $\nu_{32} = -3.0$, $\nu_{32} = 0.0$, and $\nu_{32} = 1.0$. \textbf{(b)}~Inverse design for unseen target property combinations, where both the stress-strain response and Poisson's ratio ($\nu_{32} = -3.0$) extend well beyond the training distribution. Each example shows the achieved Poisson's ratio of the best match from the training dataset and the generated designs, compared to the distribution of training samples in the relevant property space (each dot represents a unit cell in the training data, color-coded by shell thickness). The effective directional Young's modulus $E_{33}$ is normalized by the Young's modulus of the base material $E_s$ and the relative density $\rho$. The simulated samples at 10\% compressive strain illustrate their lateral deformations. All shown stress-strain responses were obtained via FE analysis, with NRMSE values (in brackets) quantifying design accuracy relative to target responses.
	
	\textbf{Figure 6}: \textbf{Experimental validation of shell metamaterials generated by DiffuMeta.} \textbf{(a)}~Fabricated representative shell samples alongside their corresponding Computer-Aided Design (CAD) models. Samples are arranged in $5 \times 5 \times 1$ arrays with $10\%$ relative density. All scale bars are \SI{10}{\milli\meter}. \textbf{(b-d)}~Comparison between experimental stress-strain responses (dashed lines) and finite element simulation results (solid lines) for three distinct shell designs generated by DiffuMeta. Target responses used as conditions for the diffusion model are indicated with red triangles. The deformed configurations of generated shell unit cells obtained from FE analysis at different strain levels are shown in $5 \times 5 \times 1$ tessellations for visualization. The experimental results demonstrate overall good agreement with both the target responses and simulation predictions for the nonlinear behavior, validating the effectiveness of the inverse design framework. 
	

	\end{document}


\begin{center}
{\Large\sffamily\bfseries Supplementary Information}\\[0.3cm]
{\large Algebraic language models for the inverse design of metamaterials via diffusion transformers}\\[0.2cm]
Li Zheng, Siddhant Kumar, and Dennis M. Kochmann
\end{center}
\vspace{1cm}

\pagenumbering{arabic}
\setcounter{page}{1}
\setcounter{tocdepth}{2}
\makeatletter
\ifdefined\@linkcolor
  \edef\SI@saved@linkcolor{\@linkcolor}%
  \hypersetup{linkcolor=black}%
  \tableofcontents
  \hypersetup{linkcolor=\SI@saved@linkcolor}%
\else
  \tableofcontents
\fi
\makeatother
\thispagestyle{empty}

\newpage
\section{Dataset generation}\label{sec:SI-dataset-generation}
\subsection{Dataset generation procedure}

To construct an extensive design space for shell metamaterials, we systematically generate implicit surface equations based on periodic nodal surfaces~\cite{gandy2001nodala} (PNS), using Fourier-type basis functions. These naturally ensure smooth, continuous, and non-intersecting geometries that are periodic in all spatial dimensions. The dataset generation process begins with the construction of a base library of trigonometric function terms, as detailed in Table~\ref{tab:SI-function-terms}. These terms are categorized into first-order (individual trigonometric functions), second-order (multiplicative combinations of two functions), and third-order (products of three functions) components. This hierarchical organization enables systematic exploration of increasingly complex surface topologies while maintaining computational tractability. For each candidate implicit surface equation, we randomly sample one to three trigonometric terms from the base library {without replacement} and combine those with randomly sampled coefficients. The coefficients $\alpha_i$ and constant offset $c$ are drawn from a uniformly spaced grid spanning $(-6, 6)$ in increments of~0.1, providing extensive parameter coverage while ensuring computational tractability {(see also Table~\ref{tab:SI-coefficient-bounds-study} for a systematic evaluation of different sampling schemes)}. The general form of our implicit equations follows as
\begin{equation}
    \Psi(x,y,z) = \sum_{i \in \mathcal{I}} \alpha_{i}T_i(x,y,z) + c = 0,
\end{equation}
where $T_i$ represents the trigonometric function terms from our base library, $\mathcal{I}\subset\{1,2,\ldots,N\}$ is a randomly selected subset with $1\leq|\mathcal{I}|\leq 3$, and $N$ is the total number of tokens in the library. {As the sampling order is random, algebraically equivalent expressions (e.g., $T_i + T_j$ and $T_j + T_i$ or $T_i \cdot T_j$ and $T_j \cdot T_i$) can occur in different instances. While such exact duplicates are statistically rare given the vast combinatorial space ($\sim 10^{12}$ possible equations), geometrically similar structures can still arise through distinct mathematical representations. We note that the chosen trigonometric functions do not form a strictly independent mathematical basis, as linear dependencies exist among certain term combinations (e.g., $\sin(2x)=2\sin(x)\cos(x)$). We intentionally preserve this redundancy, as it allows the model to encounter and learn from multiple equivalent or near-equivalent representations of similar geometries, ultimately improving both robustness and generalization performance. In this regard, the natural redundancy in our expression space serves as an {implicit form of data augmentation} that emerges from the mathematical structure of our design representation.}

\begin{figure}[!htbp]
    \centering
    \includegraphics[width = 0.95\textwidth]{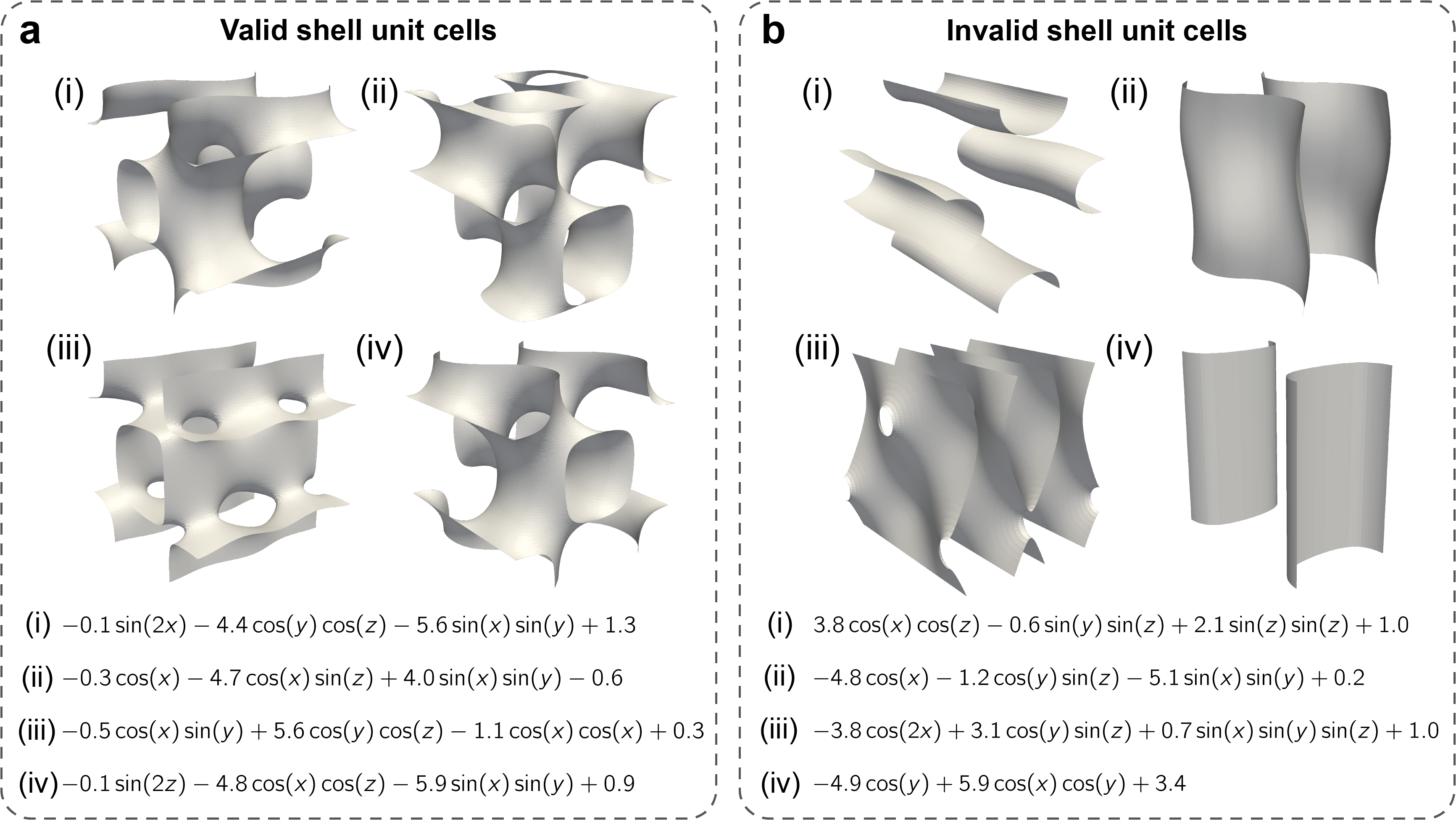}
    \captionsetup{justification=justified}
    \caption{\textbf{Representative examples of shell designs generated by our dataset generation process.} Examples of \textbf{(a)}~valid shell designs and \textbf{(b)}~invalid shell designs with disjoint components or non-periodic structures. Also shown are the corresponding implicit surface equations.}
    \label{fig:SI-shell-designs}
\end{figure}

\begin{table}[!ht]
    \centering
    \caption{\textbf{Base library of trigonometric functions used for constructing implicit surface equations.}}
    \begin{tabular}{ll}
    \toprule
    \textbf{Categories} & \textbf{Function Terms} \\ 
    \midrule
    First-order & $\cos(x)$, $\cos(y)$, $\cos(z)$, $\sin(x)$, $\sin(y)$, $\sin(z)$, \\
               & $\cos(2x)$, $\cos(2y)$, $\cos(2z)$, $\sin(2x)$, $\sin(2y)$, $\sin(2z)$ \\
    \midrule
    Second-order & $\cos(x)\cos(y)$, $\cos(x)\sin(y)$, $\cos(x)\cos(z)$, $\cos(x)\sin(z)$, \\
                & $\cos(y)\cos(z)$, $\cos(y)\sin(z)$, $\sin(x)\cos(y)$, $\sin(x)\sin(y)$, \\
                & $\sin(y)\sin(z)$, $\sin(y)\cos(z)$, $\sin(x)\cos(z)$, $\sin(x)\sin(z)$, \\
                & $\cos^2(x)$, $\cos^2(y)$, $\cos^2(z)$, $\sin^2(x)$, $\sin^2(y)$, $\sin^2(z)$ \\
    \midrule
    Third-order & $\cos(x)\cos(y)\cos(z)$, $\sin(x)\sin(y)\sin(z)$ \\
    \bottomrule
    \end{tabular}
    \captionsetup{justification=justified}
    \label{tab:SI-function-terms}
\end{table}
\newpage
To ensure physical feasibility, we implement validation criteria for each generated equation: (1) the resulting implicit function must form a self-connected surface in 3D space, (2) the surface must retain periodicity in all spatial dimensions. To validate the geometric feasibility of each generated surface, we discretize the implicit function onto a regular 3D grid with $70 \times 70 \times 70$ resolution and employ the Marching Cubes algorithm~\cite{lorensen1998marching} to extract the zero-level set corresponding to $\Psi(x,y,z) = 0$. The extracted triangulated surface mesh is then treated as a graph, for which we verify that it contains exactly one connected component to ensure structural continuity. Periodicity is verified by ensuring the selected trigonometric terms collectively depend on all three spatial variables $x$, $y$, and $z$, which automatically preserves the overall periodicity of $2\pi$ in all directions, since all base library terms have periods of $2\pi$ or $\pi$. Invalid surfaces (e.g., those with disjoint components or non-periodic structures) are discarded during the generation process. The complete dataset generation procedure is summarized in Algorithm~\ref{alg:dataset-generation}. Representative examples of valid and invalid shell designs generated with this approach and their corresponding implicit surface equations are shown in Fig.~\ref{fig:SI-shell-designs}.

\begin{algorithm}[!htbp]
    \caption{Shell metamaterial dataset generation}
    \label{alg:dataset-generation}
    \begin{algorithmic}[1]
    \Require Base library of trigonometric terms $\mathcal{T} = \{T_1, T_2, \ldots, T_N\}$
    \Require Coefficient range $(\alpha_{\min}, \alpha_{\max}) = (-6, 6)$ with step size $\Delta\alpha = 0.1$
    \Require Target dataset size $M$
    \Ensure Dataset $\mathcal{D} = \{(\Psi_k, \boldsymbol{\sigma}_k, \mathbb{C}_k)\}_{i=1}^{|\mathcal{D}|}$ of valid shell designs
    
    \State Initialize empty dataset $\mathcal{D} \leftarrow \emptyset$
    \State $k \leftarrow 0$ \Comment{Counter for generated designs}
    
    \While{$|\mathcal{D}| < M$}
        \State $k \leftarrow k + 1$ \Comment{Sample equation structure}
        \State $n_{\text{terms}} \leftarrow $ RandomInteger$(1, 3)$ \Comment{Select number of trigonometric terms}
        \State $\mathcal{I} \leftarrow $ RandomSample$(\{1, 2, \ldots, N\}, n_{\text{terms}})$ \Comment{Select term indices}
        \For{$i \in \mathcal{I}$}
            \State $\alpha_i \leftarrow $ RandomChoice$((\alpha_{\min} : \Delta\alpha : \alpha_{\max}))$ \Comment{Sample coefficients}
        \EndFor
        \State $c \leftarrow $ RandomChoice$((\alpha_{\min} : \Delta\alpha : \alpha_{\max}))$ \Comment{Sample constant offset}
        \State $\Psi_k(x,y,z) \leftarrow \sum_{i \in \mathcal{I}} \alpha_i T_i(x,y,z) + c$ \Comment{Construct implicit equation}
        
        \State Extract isosurface $S_k$ at $\Psi_k = 0$ \Comment{Extract and validate surface geometry}
        \If{$S_k$ is periodic in all spatial dimensions \textbf{and} $(S_k)$ forms a self-connected surface}
            \State Generate 3D geometry $G_k$ from validated surface $S_k$
            \State Calculate the surface area $A_k$ and scale wall thickness by $h = \rho_{\text{target}}\cdot V_{\text{cube}}/{A_k}$
            \State Conduct FE simulation to obtain stress-strain responses  $\boldsymbol{\sigma}_k$ and homogenized stiffness tensor $\mathbb{C}_k$
            \State $\mathcal{D} \leftarrow \mathcal{D} \cup \{(\Psi_k, \boldsymbol{\sigma}_k, \mathbb{C}_k)\}$
        \EndIf
    \EndWhile
    \State \Return $\mathcal{D}$
    \end{algorithmic}
\end{algorithm}  

\subsection{Generalizability of the parameterization}
{
Despite the apparent simplicity of using only trigonometric functions, the current framework provides substantial design flexibility, creating a theoretical design space of approximately $10^{12}$ possible unique implicit equations. While our dataset of 23,534 samples represents only a small fraction of this theoretical space, our analysis demonstrates that it covers a remarkably wide range of mechanical behaviors. This is evidenced, e.g., by the diverse stress-strain responses (Fig.~1c, main article) and Poisson's ratios (Fig.~5a, main article) achieved in our dataset, showing that the selected samples span a rich property space. Thus, the chosen function set provides a computationally tractable yet expressive parameterization for exploring complex designs. Note that this design space is many orders of magnitude more complex than those currently studied in related literature on triply-periodic surfaces~\cite{wang2022ihgan,wang2022inversea,perroni-scharf2025dataefficientb,jadhav2024generative,liu2024inverse} (which classically consider only a very small subset of empirically chosen topologies).

Importantly, our equation-as-sequence approach is inherently generalizable and not restricted to trigonometric functions. Additional mathematical operations, such as exponential functions, logarithmic functions, polynomial terms, or even more complex operations like hyperbolic functions, can be incorporated by treating them as new tokens in the vocabulary. The main challenge, however, lies in ensuring that these new functions still yield geometrically valid and connected surfaces. Our goal is to formulate the design problem in an expression space that naturally favors geometrically feasible and physically meaningful structures, which are further easy to tessellate in 3D. In this regard, trigonometric basis functions are particularly advantageous due to their inherent periodicity and smoothness properties, which tend to promote continuous and manufacturable level-set surfaces. However, the framework itself is general and readily adaptable to broader function sets, offering greater flexibility than traditional optimization approaches that rely on fixed function templates.

To demonstrate this flexibility, we systematically investigated how different coefficient bounds and discretization increments affect the constructed design space, as summarized in Table~\ref{tab:SI-coefficient-bounds-study}. The validity rate represents the percentage of randomly sampled equations that yield geometrically valid and self-connected surfaces suitable for mechanical testing. As shown in Table~\ref{tab:SI-coefficient-bounds-study}, expanding the coefficient bounds or refining the discretization exponentially increases the theoretical design space (by one to two orders of magnitude) without significantly improving the validity rate. In fact, finer discretization and larger bounds slightly reduce the validity rate. The selected parameterization (coefficient bounds of $(-6, 6)$ and increment of $0.1$) therefore offers a practical compromise between coverage of the design space and computational efficiency for dataset generation and model training.

\begin{table}[h!]
\centering
\caption{{\textbf{Effect of coefficient bounds and discretization increments on the design space.} Validity rates are averaged over multiple sampling trials (1,000, 2,000, 3,000, and 5,000 equations per trial).}}
\label{tab:SI-coefficient-bounds-study}
\begin{tabular}{cccc}
\toprule
{Coefficient Bounds} & {Increment} & {Theoretical Design Space Size} & {Validity Rate (\%)} \\
\midrule
$\boldsymbol{(-6,\,6)}$ & \textbf{0.1} & $\boldsymbol{\sim 10^{12}}$ & \textbf{3.2} \\
$(-9, 9)$ & 0.1 & $\sim 10^{13}$ & 3.2 \\
$(-6, 6)$ & 0.05 & $\sim 10^{13}$ & 3.0 \\
$(-9, 9)$ & 0.05 & $\sim 10^{14}$ & 2.6 \\
\bottomrule
\end{tabular}
\end{table}

}
\section{FE simulation}
\label{sec:SI-FE-simulation}

\begin{figure}[!hb]
    \centering
    \includegraphics[width = \textwidth]{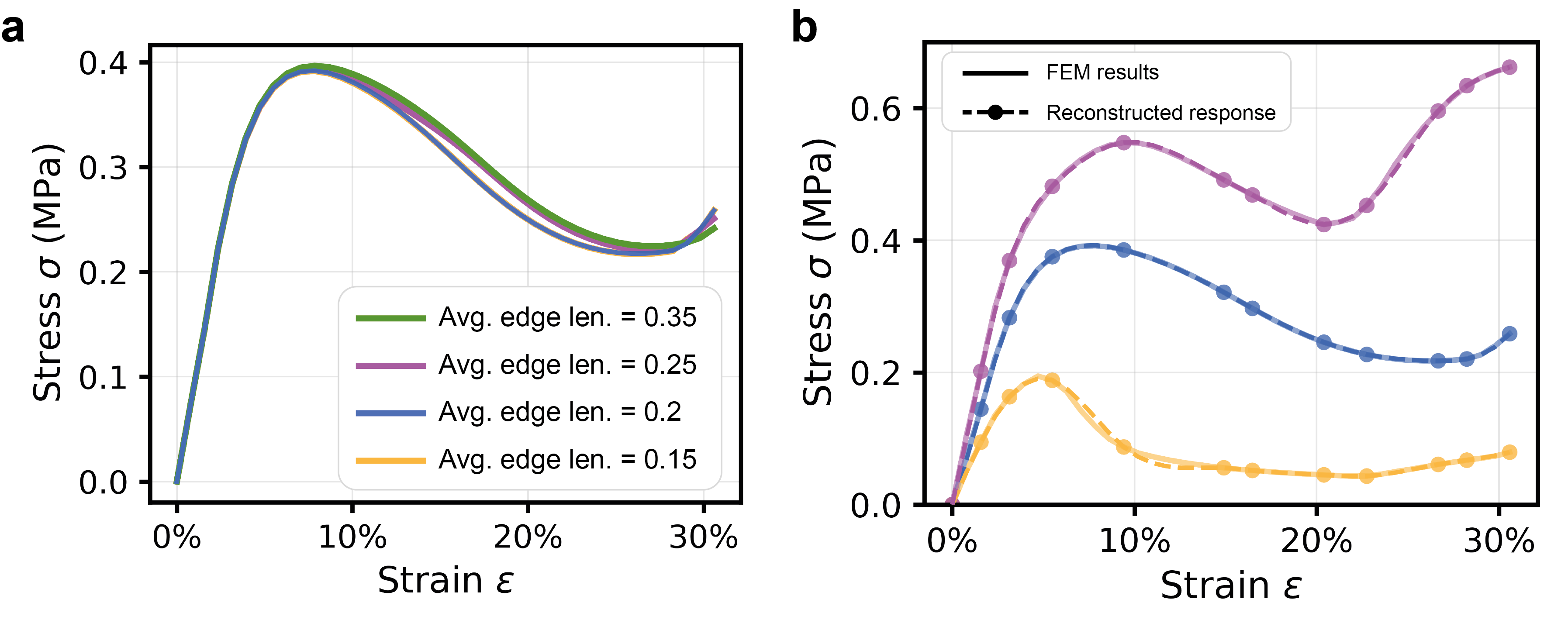}
    \captionsetup{justification=justified}
    \caption{\textbf{FE simulation results.} \textbf{(a)}~Mesh refinement convergence study for a representative shell unit cell (shown in Fig.~1 of the main article) with dimensions $10 \times 10 \times 10~\mathrm{mm}$  loaded in uniaxial compression. Three-node triangular shell elements (S3R) are used for the simulation. Meshes are generated with an average edge length of 0.35, 0.25, 0.2, and 0.15~$\mathrm{mm}$. \textbf{(b)}~Comparison between the original stress-strain curve (obtained via FE simulation, solid line) and the reconstructed curve (dashed line)  using cubic spline interpolation of the 11 sampled stress values (dots).}
    \label{fig:SI-mesh-convergence}
\end{figure}

To assess the accuracy of FE simulations, we conduct a mesh convergence study. For this purpose, we select the representative shell structure in Fig.~1 of the main article (with dimensions $10 \times 10 \times 10~\mathrm{mm}$) and uniformly refine the mesh. We use three-node triangular shell elements (S3R) and simulate the compressive stress-strain responses in ABAQUS/CAE 2023. Convergence was assessed by monitoring the relative change in the computed response. For each mesh refinement level $i$, we calculated the relative error as
\begin{equation}
    \epsilon_i = \frac{\|\boldsymbol{\sigma}_i - \boldsymbol{\sigma}_{i-1}\|^2}{\|\boldsymbol{\sigma}_{i-1}\|^2} \times 100\%,
\end{equation}
where $\boldsymbol{\sigma}_i$ represents the stress values computed with mesh refinement level $i$. The convergence analysis reveals that the relative errors decrease monotonically with mesh refinement, as shown in Fig.~\ref{fig:SI-mesh-convergence}a. For average mesh edge lengths of 0.35, 0.25, 0.2, and 0.15~$\mathrm{mm}$, the corresponding relative errors are 3\%, 1.43\%, and 0.26\%, respectively. We established a convergence criterion of less than 1\% relative change in stress between consecutive mesh refinements. Based on this convergence study, we selected a mesh density corresponding to an average edge length of 0.2~$\mathrm{mm}$ for all simulations in our dataset generation process. This choice ensures computational accuracy (relative error $< 0.3$\%) while incurring reasonable computational costs for the large-scale dataset generation.

To facilitate efficient machine learning model training, we represent each complete stress-strain curve as a reduced-dimensional vector $\boldsymbol{\sigma}\in\mathbb{R}^{11}$ by sampling the compressive stress values at 11 strategically chosen strain levels that span the full applied strain range, as shown in Table~\ref{tab:strain-points}. These sampling points are selected to capture the key features of the nonlinear stress-strain curve, including the initial elastic regime, onset of nonlinearity, and post-yield and/or post-buckling response. Note that the initial data point at zero strain and zero stress ($\varepsilon = 0$, $\sigma = 0$) is tacitly assumed and excluded from the sampling for computational efficiency. To validate this dimensionality reduction approach, we demonstrate that the original full stress-strain curves can be accurately reconstructed using cubic spline interpolation of the 11 sampled stress values (achieving reconstruction NRMSE values below 0.6\% for $2,000$ randomly sampled data points from our dataset), as illustrated in Fig.~\ref{fig:SI-mesh-convergence}b. 

\begin{table}[!hb]
    \centering
    \caption{\textbf{Strain evaluation points used for stress-strain curve sampling.} The 11 strain levels are selected to capture the key features of the stress-strain responses. The initial step at $0\%$ strain is excluded for efficiency.}
    \label{tab:strain-points}
    \begin{tabular}{c|c|c|c|c|c|c|c|c|c|c|c}
        \hline
        \textbf{Index} & 1 & 2 & 3 & 4 & 5 & 6 & 7 & 8 & 9 & 10 & 11 \\
        \hline
        \textbf{Strain (\%)} & 1.57 & {3.14} & 5.49 & 9.41 & 14.9 & 16.5 & 20.4 & 22.7 & 26.7 & {28.2} & 30.0 \\
        \hline
    \end{tabular}
\end{table}

\section{Diffusion model}\label{sec:SI-diffusion-model}
\subsection{Continuous diffusion models}
Diffusion models are a class of generative models that learn to reverse a diffusion process to generate samples from complex distributions~\cite{ho2020denoising}. The core of diffusion models involves two processes: a forward process that gradually corrupts data by adding noise, and a reverse process that learns to recover the original data by progressively denoising. 
The forward diffusion process systematically corrupts the original data by progressively adding Gaussian noise over $T$ discrete time steps. This process forms a Markov chain where each step depends only on the previous state as
\begin{equation}
    q(\bfx_t| \bfx_{t-1}) = \mathcal{N}(\bfx_t; \sqrt{1-\beta_t}\bfx_{t-1}, \beta_t \bfI),
\end{equation}
where $\{\beta_t \in (0,1)\}_{t=1}^T$ is a sequence of hyperparameters controlling the noise variance at each diffusion step $t$. We can then directly sample from any intermediate time step without sequential computation via
\begin{equation}
    q(\bfx_t| \bfx_0) = \mathcal{N}(\bfx_t; \sqrt{\Bar{\alpha}_t}\bfx_0, (1-\Bar{\alpha}_t)\bfI), \quad \text{with} \quad \alpha_t = 1 - \beta_t, \Bar{\alpha}_t = \prod_{i=1}^t \alpha_i.
\end{equation}
This allows us to express any noisy sample $\bfx_t$ as
\begin{equation}
    \bfx_t = \sqrt{\Bar{\alpha}_t}\bfx_0 + \sqrt{1-\Bar{\alpha}_t}\boldsymbol{\varepsilon},
\end{equation}
where $\boldsymbol{\varepsilon} \sim \mathcal{N}(\bfnull, \bfI)$ represents the accumulated noise.

The reverse process aims to reconstruct the original data by learning to remove noise progressively, which can be modeled by a neural network $p_{\theta}$ parameterized by $\theta$ such that
\begin{equation}
    p_{\theta}(\bfx_{t-1} | \bfx_t) = \mathcal{N}(\bfx_{t-1}; \mu_{\theta}(\bfx_t, t), \Sigma_{\theta}(\bfx_t, t)),
\end{equation}
where $\mu_{\theta}(\bfx_t, t)$ and $\Sigma_{\theta}(\bfx_t, t)$ represent the predicted mean and covariance. The marginal likelihood of generating the original data is then given by
\begin{equation}\label{SI-eq:p-theta}
    p_{\theta}(\bfx_0) = \int p_{\theta}(\bfx_{0:T})d\bfx_{1:T}.
\end{equation}
Since the forward process is fixed, only the reverse process contains trainable parameters. However, maximization of the (log-)likelihood $\log p_{\theta}$ in Eq.~\ref{SI-eq:p-theta} is computationally intractable. We instead maximize the tractable variational lower bound~\cite{kingma2013auto} on the log-likelihood, leveraging Jensen's inequality~\cite{ho2020denoising}:
\begin{equation}
    \begin{aligned}
        \log p_{\theta}(\bfx_0) & = \log \int p_{\theta}(\bfx_{0:T})d\bfx_{1:T} \\
        & \geq \mathbb{E}_{q(\bfx_{1:T}|\bfx_0)} \left[ \log \dfrac{p_{\theta}(\bfx_{0:T})}{q(\bfx_{1:T}|\bfx_0)} \right] \\
        & = \mathbb{E}_{q(\bfx_{1:T}|\bfx_0)} \left[ \log \dfrac{p(\bfx_T)\prod^T_{t=1} p_{\theta}(\bfx_{t-1}|\bfx_t)}{\prod^T_{t=1}q(\bfx_t|\bfx_{t-1})} \right].
    \end{aligned}
\end{equation}
Based on the above, we can train a model to learn the reverse process and reconstruct the data by optimizing the variational lower bound of the log-likelihood for $p_{\theta}(\bfx_0)$, which reads
\begin{equation}\label{eq:loss-vlb-x0}
    \scalebox{0.95}{$
    \begin{aligned}
        \mathcal{L}_{\text{vlb}}^{\theta}(\bfx_0) &= \underbrace{-\mathbb{E}_{q(\bfx_1|\bfx_0)} \left[ \log p_{\theta}(\bfx_0 | \bfx_1) \right]}_{\mathcal{L}_0} + \sum_{t=2}^{T}\underbrace{\mathbb{E}_{q(\bfx_t|\bfx_0)} \left[\mathcal{D}_{\text{KL}}(q(\bfx_{t-1}|\bfx_t, \bfx_0) \parallel p_{\theta}(\bfx_{t-1} | \bfx_t)) \right]}_{\mathcal{L}_{t-1}} \\
        & \quad + \underbrace{\vphantom{-\mathbb{E}_{q(\bfx_1|\bfx_0)}}\mathcal{D}_{\text{KL}}(q(\bfx_T|\bfx_0) \parallel p(\bfx_T))}_{\mathcal{L}_T}
    \end{aligned}
    $},
\end{equation}
where $\mathcal{D}_{\text{KL}}(a \| b)$ denotes the Kullback-Leibler (KL) divergence between two distributions $a$ and $b$. The second term, which is the KL-divergence between two multivariate Gaussian distributions, has a closed-form solution:
\begin{equation}
    \mathbb{E}_{q(\bfx_t|\bfx_0)} \left[\mathcal{D}_{\text{KL}}(q(\bfx_{t-1}|\bfx_t, \bfx_0) \parallel p_{\theta}(\bfx_{t-1} | \bfx_t)) \right] = \mathbb{E}_{q(\bfx_t|\bfx_0)} \left[ \dfrac{1}{2 \sigma_t^2} \|\mu_{\theta}(\bfx_t, t) - \hat{\mu}(\bfx_t, \bfx_0) \|^2 \right]+C,
\end{equation}
where $C$ is a constant, $\mu_{\theta}(\bfx_t, t)$ is the mean of $p_{\theta}(\bfx_{t-1} | \bfx_t)$ predicted by the model, and $\hat{\mu}(\bfx_t, \bfx_0)$ is the mean of the posterior distribution $q(\bfx_{t-1} | \bfx_t, \bfx_0)$. This yields a loss term
$\mathcal{L}_{\text{simple}}(\theta)\coloneqq  \mathbb{E}_{q(\bfx_{1:T}|\bfx_0)} \left[ \|\mu_{\theta}(\bfx_t, t) - \hat{\mu}(\bfx_t, \bfx_0) \|^2 \right]$. 

The training objective can be reparameterized to train a model $\boldsymbol{\varepsilon}_{\theta}(\bfx_t, t)$ to directly predict the noise $\boldsymbol{\varepsilon}$ that was added during the forward process, which simplifies the optimization and improves training stability~\cite{ho2020denoising}. This reparameterization leads to the simplified loss function
\begin{equation}\label{eq:loss-simple-noise}
    \theta \leftarrow \argminWithArgs_{\theta}\,\mathcal{L}(\bfx_0) = \argminWithArgs_{\theta}\,\mathbb{E}_{\bfx_0, \boldsymbol{\varepsilon}, t} \left[\lVert \boldsymbol{\varepsilon} - \boldsymbol{\varepsilon}_{\theta}(\bfx_t, t) \rVert^2 \right],
\end{equation}
where $\bfx_t = \sqrt{\Bar{\alpha}_t}\bfx_0 + \sqrt{1-\Bar{\alpha}_t}\boldsymbol{\varepsilon}$ with $\boldsymbol{\varepsilon} \sim \mathcal{N}(\bfnull, \mathbf{I})$. In this work, we train the denoising network $f_{\theta}(\bfx_t, t)$ to directly predict the clean inputs $\bfx_0$ from the noisy inputs $\bfx_t$, which aligns well with the word embedding space for text generation~\cite{li2022diffusionlm}. The loss term in $\mathcal{L}_{\text{simple}}(\theta)$ can be simplified as
\begin{equation}\label{eq:mean-q}
    \scalebox{0.95}{$
    \begin{aligned}
    \lVert \mu_{\theta}(\bfx_t, t) - \hat{\mu}(\bfx_t, \bfx_0) \rVert^2 & = \lVert (\dfrac{\sqrt{\alpha_t}(1-\Bar{\alpha}_{t-1})}{1-\Bar{\alpha}_t}\bfx_t + \dfrac{\sqrt{\Bar{\alpha}_{t-1}}\beta_t}{1-\Bar{\alpha}_t}\bfx_0) - (\dfrac{\sqrt{\alpha_t}(1-\Bar{\alpha}_{t-1})}{1-\Bar{\alpha}_t}\bfx_t + \dfrac{\sqrt{\Bar{\alpha}_{t-1}}\beta_t}{1-\Bar{\alpha}_t}f_{\theta}(\bfx_t,t)) \rVert^2 \\
    &= \lVert \dfrac{\sqrt{\Bar{\alpha}_{t-1}}\beta_t}{1-\Bar{\alpha}_t} (\bfx_0 - f_{\theta}(\bfx_t,t))\rVert^2\\
    & \propto \lVert \bfx_0 - f_{\theta}(\bfx_t,t)\rVert^2.
\end{aligned}
$}
\end{equation}
\subsection{End-to-end training of diffusion language models}

To extend the continuous diffusion models to the text domain, we adopt the Diffusion-LM~\cite{li2022diffusionlm} framework, which maps discrete text into a continuous latent space via an embedding function $\bfe_{\phi}$. In our approach, we represent each shell geometry by its implicit equation, which we treat as a discrete sequence of mathematical tokens: $\bfw = [w_1, w_2, \ldots, w_n]$, where $n$ denotes the sequence length. Individual tokens $w_i$ represent mathematical components such as trigonometric functions, arithmetic operators, coefficients, or variables. To enable continuous diffusion modeling, we map this discrete tokenized sequence into a continuous embedding space through a learnable embedding function $\bfe_{\phi}$:
\begin{equation}
    \bfe_{\phi}(\bfw)=[\bfe_{\phi}(w_1), \bfe_{\phi}(w_2), \ldots, \bfe_{\phi}(w_n)]^T \in \mathbb{R}^{n \times d},
\end{equation}
where $d$ is a hyperparameter denoting the embedding dimension. Therefore, the original forward process is extended to a new Markov transition defined by $q_{\phi}(\bfx_0 \mid \bfw) = \mathcal{N}(\bfe_{\phi}(\bfw), \sigma_0\bfI)$. 

In the reverse process, the goal is to recover the original input embeddings through iterative denoising, which are subsequently mapped back to the discrete sequence $\bfw$ representing the implicit equation. This is achieved via a trainable rounding step, parameterized by $p_{\theta}(\bfw\mid\bfx_0) = \prod_{i = 1}^n p_{\theta}(w_i \mid x_i)$, where each $p_{\theta}(w_i \mid x_i)$ is computed as a softmax over the token vocabulary. During decoding, the most probable token is selected according to $\argmax\,\,p_{\theta}(\bfw \mid \bfx_0)$ for each position, thus reconstructing the discrete sequence from the denoised embeddings $\bfx_0$. In this work, our framework is trained to jointly learn the diffusion model's parameters and word embeddings. The training objective in Eq.~\ref{eq:loss-vlb-x0} can be extended to
\begin{equation}
    \begin{aligned}
    \mathcal{L}_{\text{vlb}}^{\phi,\theta}(\bfw) &= \mathbb{E}_{q_{\phi}(\bfx_0|\bfw)} \left[
        \mathcal{L}_{\text{vlb}}(\bfx_0) + \log q_{\phi}(\bfx_0\mid\bfw) - \log p_{\theta}(\bfw\mid\bfx_0)
    \right]\\
    &= \mathbb{E}_{q_{\phi}(\bfx_0|\bfw)} \left[
        \underbrace{\log\dfrac{q(\bfx_T\mid\bfx_0)}{p_{\theta}(\bfx_T)}}_{L_T} + \sum_{t=2}^{T} \underbrace{\log\dfrac{q(\bfx_{t-1}\mid\bfx_0,\bfx_t)}{p_{\theta}(\bfx_{t-1}\mid\bfx_t)}}_{L_{t-1}} - \underbrace{\dfrac{\log q_{\phi}(\bfx_0\mid \bfw)}{\log p_{\theta}(\bfx_0\mid\bfx_1)}}_{L_0} - \underbrace{\log p_{\theta}(\bfw\mid\bfx_0)}_{L_{\text{round}}}
    \right],
    \end{aligned}
\end{equation}
where $q_{\phi}(\bfx_0\mid\bfw)=\mathcal{N}(\bfe_{\phi}(\bfw), \sigma_0\bfI)$ denotes the embedding transition from discrete text $\bfw$ to continuous latent space $\bfx_0$, and $p_{\theta}(\bfw\mid\bfx_0)=\prod_{i=1}^n p_{\theta}(\bfw_i\mid\bfx_i)$ represents the rounding transition from $\bfx_0$ to $\bfw$. Applying the same simplification and reparameterization strategy as in Eq.~\ref{eq:loss-simple-noise}, we can further simplify the training objective, leading to
\begin{equation}
    \scalebox{0.85}{
    $\phi,\theta \leftarrow \argminWithArgs_{\phi,\theta}\,\mathcal{L}(\bfw) = \argminWithArgs_{\phi,\theta}\,\mathbb{E}_{q_{\phi}(\bfx_{0:T}|\bfw)} \left[
            \sum_{t=1}^T \underbrace{\lVert f_{\theta}(\bfx_t, t) - \bfx_0 \rVert^2}_{\mathcal{L}_t}  + \underbrace{\lVert \bfe_{\phi}(\bfw) - f_{\theta}(\bfx_1, 1) \rVert^2}_{\mathcal{L}_0} \underbrace{-\log p_{\theta}(\bfw|\bfx_0)}_{\mathcal{L}_{\text{round}}}\right].$
            }
\end{equation}

{
\subsection{Model architecture}
\label{subsec:SI-model-architecture}

In our framework, we chose a diffusion transformer~\cite{peebles2023scalablea} (DiT) architecture as the backbone denoising network. This selection was motivated by several key considerations specific to our problem formulation. Our equation-as-sequence parameterization represents each implicit equation as an ordered sequence of mathematical tokens, where even small changes in one term can significantly alter the resulting geometry and its mechanical behavior. Capturing such global, nonlinear dependencies across all terms requires a model capable of modeling long-range and bidirectional relationships. Transformers~\cite{vaswani2017attention}, through their self-attention mechanisms, are ideally suited for this purpose~\cite{islam2023comprehensive,fournier2023practical}. In contrast, alternative sequential models such as recurrent neural networks (RNNs) or long short-term memory (LSTM) models, process data sequentially and are less effective at capturing bidirectional dependencies~\cite{graves2005framewise}. Besides, prior work on discrete sequence generation has demonstrated the advantages of transformer-based architectures for diffusion models over U-Net structures with 1D-convolutional layers for sequence generation tasks~\cite{li2022diffusionlm}. Therefore, we adopt a diffusion transformer architecture following the Diffusion-LM framework~\cite{li2022diffusionlm}. By modeling all pairwise dependencies simultaneously through self-attention, transformers enable the network to learn robust and expressive representations of the implicit equation design space. In addition, the DiT framework provides effective conditioning mechanisms for integrating target mechanical properties (e.g., stress-strain responses and effective elastic properties) into the generative process via cross-attention and adaptive layer normalization (adaLN) blocks~\cite{peebles2023scalablea}. This enables the model to generate shell designs that not only remain structurally valid but also align with prescribed mechanical targets.

In summary, the diffusion transformer architecture is particularly well-suited to our formulation, as it combines the sequential modeling capabilities of transformers with the generative stability and diversity of diffusion models, effectively addressing the challenges of discrete sequence generation, complex design constraints, and conditional property-driven generation within a unified framework.
}

{
\section{Ablation studies}
\label{sec:SI-ablation-studies}

\subsection{Effect of dataset size}

In this section, we provide a systematic ablation study to assess the effect of dataset size on the performance of our diffusion transformer model. The dataset size of 23,534 samples in our work was determined by balancing the constraints of computational feasibility and model performance. As reported in Table~\ref{tab:SI-runtime}, each finite element simulation requires approximately 5 minutes on a 12-core Intel Xeon processor, and generating the stress-strain responses for 23,534 samples required approximately 10 days of continuous computation. To validate that this dataset size is sufficient, we conducted ablation experiments by training our diffusion transformer model on progressively smaller subsets of the full dataset (75\%, 50\%, 25\%, and 10\%), using the same training protocol. As shown in Fig.~\ref{fig:SI-dataset-size}, we evaluated both training and validation loss and conditional generation performance across these different dataset sizes.

\begin{figure}[h]
\centering
\includegraphics[width=\textwidth]{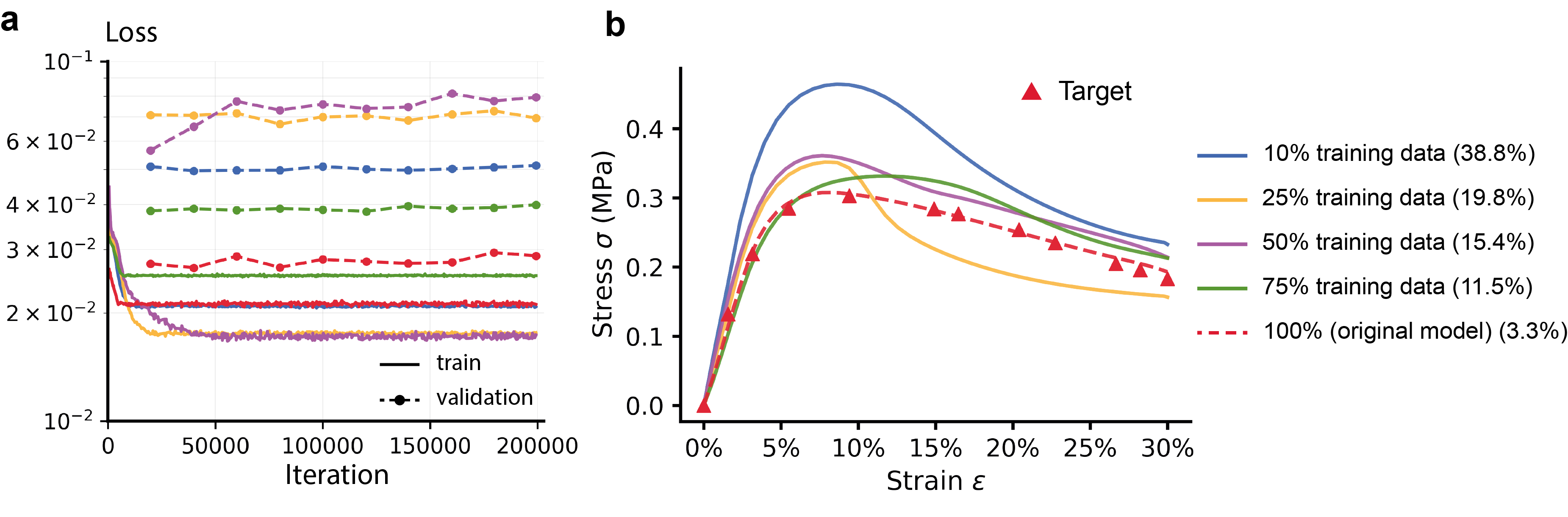}
\caption{{\textbf{Study of the effect of dataset size on the performance of the diffusion model.} \textbf{(a)} Training (solid) and validation (dotted dashed) loss curves for models trained on different fractions of the full dataset (10\%, 25\%, 50\%, 75\%, and 100\%). \textbf{(b)} Inverse design results for the same target response (corresponding to Fig.~3a in the main article), using models trained on different dataset sizes. The full dataset (100\%, red curves) achieves the lowest validation loss and NRMSE for generation and demonstrates stable convergence, confirming that the dataset size is sufficient for robust model training. (Percentages in parentheses indicate the NRMSE values.)}}
\label{fig:SI-dataset-size}
\end{figure}

Fig.~\ref{fig:SI-dataset-size}(a) demonstrates that model performance improves consistently with increasing dataset size. The full dataset achieves the lowest validation loss ($\sim$0.03) with clear convergence, while smaller datasets exhibit notably higher validation losses, indicating insufficient coverage of the design space. Although the validation loss remains higher than the training loss across all experiments (which is expected, given the large number of model parameters relative to data points), the loss curves converge smoothly, indicating stable training dynamics.

It is also important to note that in diffusion models, the training loss alone does not fully capture generation quality, as diffusion models can plateau in loss early before improvements in generation become visible~\cite{ho2020denoising,song2021scorebaseda,nichol2021improved}. Therefore, to more directly assess generation performance, we conditioned each model (trained on 10\%, 25\%, 50\%, 75\%, and 100\% of the data) on the same target stress-strain response (corresponding to Fig.~3(a) in the main article) and compared the best of 200 generated designs from each model in terms of NRMSE between the FE-evaluated and target responses. As shown in Fig.~\ref{fig:SI-dataset-size}(b), the model trained on the full dataset produces designs with superior property matching (with an NRMSE of 3.3\%) compared to models trained on smaller subsets (NRMSE ranging from 11.5\% to 38.8\%), demonstrating that the current dataset size enables robust inverse design performance.

\subsection{Effect of hyperparameters}

To determine the optimal model configuration, we performed systematic hyperparameter studies, using the validation loss as the selection criterion. Each hyperparameter was optimized sequentially, while keeping the others fixed at reasonable defaults, balancing computational efficiency with systematic exploration. For computational efficiency, these studies were conducted on 50\% of the full dataset and trained for the full 200,000 training iterations.

\begin{figure}[h]
    \centering
    \includegraphics[width=\textwidth]{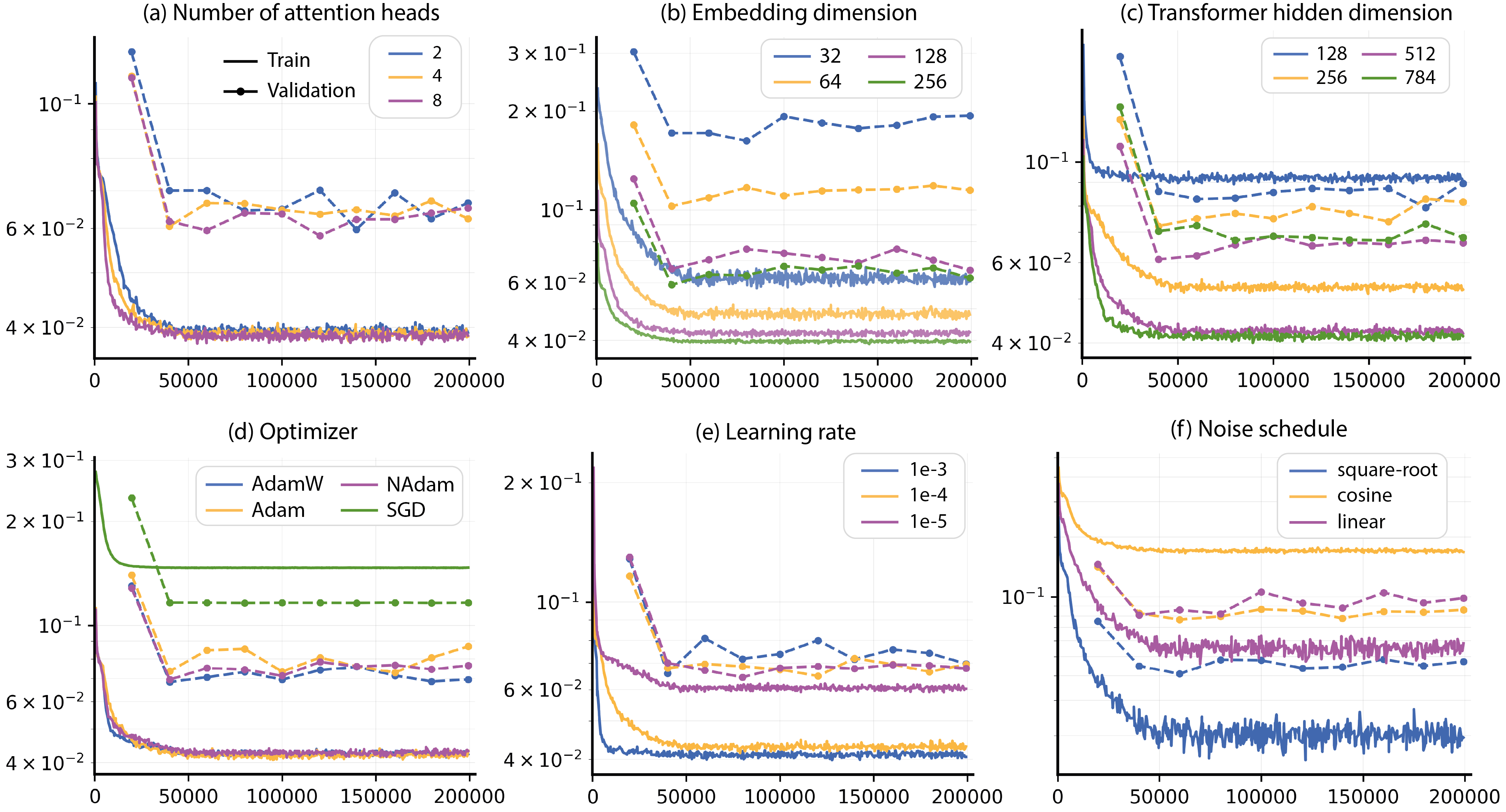}
    \caption{{\textbf{Effect of hyperparameters on model performance.} Training (solid) and validation (dotted dashed) loss curves for different hyperparameter configurations, including \textbf{(a)} number of attention heads, \textbf{(b)} token embedding dimension, \textbf{(c)} transformer hidden dimension, \textbf{(d)} optimizer, \textbf{(e)} learning rate, and \textbf{(f)} noise schedule.}} 
    \label{fig:SI-hyperparameters}
\end{figure}

\textbf{Architecture hyperparameters:} For the diffusion transformer architectural design, we explored $\{2, 4, 8\}$ attention heads, token embedding dimensions of $d \in \{32, 64, 128, 256\}$, and transformer hidden dimensions of $\{128, 256, 512, 784\}$. As shown in Fig.~\ref{fig:SI-hyperparameters}(a), 4 attention heads provided the optimal configuration, balancing the model's ability to capture diverse token relationships and model complexity. For the token embedding dimension (Fig.~\ref{fig:SI-hyperparameters}(b)), $d=128$ achieved the lowest validation loss, with smaller dimensions (32, 64) showing higher losses due to insufficient representational capacity, while the larger dimension (256) showed only marginal improvement at significantly higher computational cost. The transformer hidden dimension study (Fig.~\ref{fig:SI-hyperparameters}(c)) demonstrated that 512 dimensions offered the best performance, with smaller dimensions (128, 256) leading to underfitting, while the larger dimension (784) showed comparable performance but at significantly higher computational cost.

\textbf{Training hyperparameters:} For training hyperparameters, we evaluated optimizers (AdamW, Adam, NAdam, SGD), learning rates of $\{10^{-5}, 10^{-4}, 10^{-3}\}$, and noise schedules (square-root, linear, cosine). Fig.~\ref{fig:SI-hyperparameters}(d) shows that AdamW achieved superior convergence behavior with the lowest final validation loss and more stable training dynamics compared to other optimizers. The learning rate study (Fig.~\ref{fig:SI-hyperparameters}(e)) revealed that $10^{-4}$ provided the optimal balance, with $10^{-5}$ converging too slowly and $10^{-3}$ exhibiting training instabilities. Finally, for the noise schedule (Fig.~\ref{fig:SI-hyperparameters}(f)), the square-root schedule following the Diffusion-LM framework~\cite{li2022diffusionlm} outperformed both linear and cosine schedules, confirming its effectiveness for discrete sequence generation tasks.

Based on these studies, the optimal architectural and training hyperparameters were selected and used for all results reported in the main manuscript. The complete set of architecture hyperparameters for the denoising diffusion model is summarized in Table~\ref{tab:SI-model-architecture}, while the training hyperparameters are provided in Table~\ref{tab:SI-training-hyperparameters}.

\begin{table}[!ht]
    \centering
    \caption{\textbf{Denoising diffusion model architecture hyperparameters.}}
    \begin{tabular}{ll}
    \toprule
    \textbf{Hyperparameter} & \textbf{Value} \\ 
    \midrule
    Sequence length & 22 \\
    Token embedding dimension & 128 \\
    Number of DiT blocks & 6 \\
    Number of attention heads & 4 \\
    Transformer hidden dimension & 512 \\
    DiT block normalization & Adaptive layer norm~\cite{peebles2023scalablea} \\
    Activation function in DiT & SiLU~\cite{elfwing2017sigmoidweighted} \\
    Dropout ratio & 0.1 \\
    \bottomrule
    \end{tabular}
    \captionsetup{justification=justified}
    \label{tab:SI-model-architecture}
\end{table}

\begin{table}[!ht]
    \centering
    \caption{\textbf{Training hyperparameters for the DiffuMeta model.}}
    \begin{tabular}{ll}
    \toprule
    \textbf{Hyperparameter} & \textbf{Value} \\ 
    \midrule
    Total dataset size & 23,534 \\
    Validation data size & 10\% of total dataset\\
    Batch size & 512 \\
    Learning rate & $10^{-4}$ \\
    Optimization algorithm & AdamW~\cite{loshchilov2019decoupled} \\
    Iterations & 200,000 \\
    Diffusion steps & 2,000 \\
    Noise schedule & square-root~\cite{li2022diffusionlm} \\
    Maximum gradient norm & 1.0 \\
    Dropout ratio (for classifier-free guidance~\cite{ho2022classifierfree}) & 0.1 \\
    \bottomrule
    \end{tabular}
    \captionsetup{justification=justified}
    \label{tab:SI-training-hyperparameters}
\end{table}
}
\newpage
\section{Computational efficiency estimates}
\label{sec:SI-computational-efficiency}
To estimate the computational efficiency of our approach, we provide an overview of the runtime required for data generation, training, and sampling using our framework in Table \ref{tab:SI-runtime}. { An important consideration in evaluating data-driven inverse design approaches is the comparison between generative modeling and direct search within an expanded dataset. While the dataset search baseline performs remarkably well for interpolative cases -- which indeed highlights the richness of our dataset -- our generative approach offers several fundamental advantages that cannot be achieved through dataset expansion alone, regardless of computational resources invested.

First, as demonstrated in Section~\ref{sec:SI-unconditional-generation}, random sampling from our design space achieves only $\sim$3.2\% validity rate, meaning that expanding the dataset through naive exploration would be highly inefficient. While more advanced hardware could reduce the computational cost for FE evaluations, the fundamental inefficiency remains: most randomly sampled equations produce invalid or disconnected geometries. More importantly, such heuristic expansion provides no guarantee of achieving the desired property targets. Moreover, even with a substantially enlarged dataset, direct search would scale linearly with dataset size, and in high-dimensional property spaces (as in our multi-objective inverse design cases) finding near-exact matches becomes increasingly unlikely without exponentially expanding the dataset.

Second, our generative model provides qualitatively different capabilities compared to a database search. Our method generates new designs not present in the dataset, effectively addressing the one-to-many mapping challenge where multiple distinct geometries can achieve the same target properties. Dataset search is fundamentally limited to retrieving existing designs, offering no diversity when multiple solutions are desired. Moreover, as demonstrated in Figs.~4 and~5 of the main article, our approach discovers designs with properties extending beyond the training distribution. Dataset expansion through random sampling cannot efficiently target such extrapolative regions without prior knowledge of which equation parameters will yield the desired properties. Moreover, the mechanical properties of shell metamaterials are governed by complex, nonlinear phenomena including large deformations, plasticity, buckling, and frictional contact. These highly nontrivial effects make exhaustive exploration of the design space, whether through direct sampling or topology optimization, computationally prohibitive. This is precisely where generative models prove particularly powerful: by learning the underlying structure-property relationships from data, they can navigate the complex design space to identify equation combinations that produce target mechanical behaviors. This extrapolation capability is fundamentally beyond the reach of dataset search methods.

Lastly, while model training represents a one-time investment (60 GPU-hours), it subsequently enables the generation of arbitrarily many candidate designs at negligible cost ($\sim$0.2 second per design on GPU). In contrast, dataset expansion scales linearly: each new design requires a full FEA simulation ($\sim$5 minutes). More critically, the generative model amortizes this cost across all future inverse design queries, whereas expanded dataset search still requires exhaustive searches with no guarantee of finding exact matches or enabling property extrapolation.

In summary, the advantage of our generative approach lies not merely in computational efficiency but in its ability to learn and leverage structure-property relationships to enable novel design generation, property extrapolation, and guided navigation of the design space, all of which remain inaccessible to dataset search methods regardless of dataset size or hardware improvements.

\begin{table}[ht]
    \centering
    \caption{\textbf{Overview of the computational runtime, the software and hardware resources required for different tasks.} The reported runtimes are rough average estimates. {$^\dagger$Computations were performed on a 2.3GHz Intel Xeon Gold 5118 processor with 96GB of DDR4 memory at 2400 MHz.} $^\P$Reported runtimes are averaged over the representative examples in the main article and may vary due to the convergence of the FE solver.  $^\S$Computations were performed on a single Nvidia RTX 4090 with CUDA 11.7. $^\star$Reported runtimes are averaged over 200 samples.}
    \begin{tabular}[t]{lccc}
    \toprule
     \textbf{Tasks} &  \textbf{Software} &  \textbf{Hardware} &  \textbf{Runtime}\\
    \midrule
     Shell dataset generation&   Python &  CPU (20 cores)$^\dagger$&  5 minutes\\
      FE computations (of a single design) &  Abaqus &   CPU (single core)$^\dagger$& 6 minutes$^\P$\\
      FE computations (of the full dataset) &   Abaqus &   CPU (12 cores) & 10 days\\
      Model training &  PyTorch in Python &  GPU$^\S$ &   60 hours\\
      Sample generation &  PyTorch in Python &   GPU$^\S$ &   0.2 seconds$^\star$ \\
    \bottomrule
    \end{tabular}
    \label{tab:SI-runtime}
\end{table}

}

\section{Further results on model generation performance}
\label{sec:SI-generation-performance}
\subsection{Sampling process across different time steps}
We provide generation examples of the DiffuMeta model in Table~\ref{tab:SI-unconditional-generation}. At the beginning of the diffusion denoising process (when $t$ is close to $1.0$), the model produces mathematically incoherent expressions with random coefficients and invalid syntax, representing pure noise in the equation space.  As the diffusion process progresses toward $t=0.0$, the generated sequences gradually transform into well-structured mathematical equations with proper syntax, meaningful coefficients, and physically interpretable trigonometric terms that can represent valid shell geometries.

\begin{table}[!ht]
    \centering
    \caption{\textbf{Diffusion model sampling process for sequence generation.} Evolution of generated equation sequences during the reverse diffusion process, showing the progressive denoising from noisy input at time $t=1.0$ to clean output at time $t=0.0$. Each row represents a timestep in the sampling process, with expressions becoming more structured and coherent as the diffusion process approaches completion.}
    \renewcommand{\arraystretch}{1.2}
         \begin{tabular}{ll}
         \Xhline{2.5\arrayrulewidth}
          Time 1.0 & \cellcolor{tablecolor1} \begin{tabular}{@{}l@{}} $\cos(x)\cos(y)\sin(z)\sin(z)\sin(2y)3\sin(x)\sin(y)\sin(z)\sin(y)\sin(z)\cos(x)\cos(y)$ \\ $8\sin(z)\sin(z)\sin(x)\sin(z)1\cos(2y)838\cos(y)\cos(z)-64$ \end{tabular} \\
         \hline
          Time 0.9 & \cellcolor{tablecolor2} $++15\cos(x)\cos(x)\sin(x)2\cos(2y)+\sin(x)\cos(y)12\sin(x)\sin(x)6225\cos(x)\sin(z)$ \\
         \hline
         Time 0.8 & \cellcolor{tablecolor3} \begin{tabular}{@{}l@{}} $+811\sin(x)\sin(x)\cos(x)\cos(y)\cos(z)\sin(y)\sin(z)2\cos(2y)\cos(y)\cos(z)$ \\ $18\sin(y)\sin(z)2\sin(y)\sin(z)\sin(x)\sin(x)431$ \end{tabular} \\
         \hline
         Time 0.7 & \cellcolor{tablecolor4} \begin{tabular}{@{}l@{}} $+3\cos(z)\cos(z)8\cos(y)\sin(z)-1*2\cos(z)$ \\ $+4\sin(y)\sin(z)7\sin(y)\sin(z)\cos(x)\sin(z)-631$ \end{tabular} \\
         \hline
          Time 0.6 & \cellcolor{tablecolor5} $+928\sin(x)0\cos(y)\cos(y)2\sin(y)\sin(y)+8*6\cos(z)\cos(z)+2.1$ \\
         \hline
         Time 0.5 & \cellcolor{tablecolor6} $+4.1\sin(x)+1.8\cos(x)\cos(y)+4.7\cos(z)\cos(z)-2.5$ \\
         \hline
         Time 0.4 & \cellcolor{tablecolor7}  $+4.3\sin(x)+5.1\sin(z)\sin(z)+4.7\cos(z)\cos(z)-2.1$ \\
         \hline
         Time 0.3 & \cellcolor{tablecolor8} $+4.3\sin(x)\cos(y)+5.3\sin(y)\cos(z)+4.7\sin(y)\sin(y)-0.5$ \\
         \hline
         Time 0.2 & \cellcolor{tablecolor9} $+4.0\sin(x)\cos(y)+5.3\sin(y)\cos(z)+4.7\cos(y)\cos(y)-0.8$ \\
         \hline
         Time 0.1 & \cellcolor{tablecolor10} $+4.0\sin(x)\cos(y)+5.3\sin(y)\cos(z)+4.7\cos(y)\cos(y)-0.8$ \\
         \hline
         Time 0.0 & \cellcolor{tablecolor11} $+4.0\sin(x)\cos(y)+5.3\sin(y)\cos(z)+4.7\cos(y)\cos(y)-0.8$ \\
         \Xhline{2.5\arrayrulewidth}
    \end{tabular}
    \captionsetup{justification=justified}
    \label{tab:SI-unconditional-generation}
\end{table}

\subsection{Unconditional generation results}
\label{sec:SI-unconditional-generation}
Table~\ref{tab:SI-unconditional-novelty} provides representative examples of shell designs generated by DiffuMeta through unconditional sampling. Each generated equation is compared with its most similar counterpart from the training dataset based on the cosine similarity between their sequence embeddings, computed as:
\begin{equation}
d(\bfe_1, \bfe_2) = \frac{\bfe_1 \cdot \bfe_2}{\|\bfe_1, \bfe_2 \|^2} \quad \in [-1,1],
\end{equation}
where $\mathbf{e}_1$ and $\mathbf{e}_2$ are the embeddings of the generated and training equations, respectively. This comparison reveals how the model explores new mathematical expressions while maintaining structural coherence. The generated equations exhibit systematic variations in coefficients, function combinations, and mathematical operators, showcasing the model's ability to explore novel regions of the design space beyond the training distribution. Notably, even minor differences between generated and training equations -- such as small coefficient variations or function substitutions -- can lead to substantially different geometric topologies and consequently distinct mechanical properties. This sensitivity underscores the fundamental challenge inherent in the inverse design of shell metamaterials, where seemingly subtle mathematical modifications can lead to significant changes in structural behavior, highlighting the generative potential of the diffusion-based approach for discovering novel metamaterial designs.

\begin{table}[!ht]
    \centering
    \renewcommand{\arraystretch}{1.1}
    \caption{\textbf{Representative examples of unconditionally generated equations by DiffuMeta.} Comparison between generated equations and their most similar counterparts from the training dataset. The generated equations exhibit systematic variations in coefficients, signs, and function combinations while maintaining mathematical coherence, demonstrating the model's ability to explore novel design spaces beyond the training distribution.}
    \begin{tabular}{lp{0.44\textwidth}p{0.44\textwidth}}
    \toprule
    &\textbf{Generated Equation} & \textbf{Closest Training Equation} \\
    \midrule
    {\footnotesize (\romannumeral 1)} & {\footnotesize $-0.1\cos(2z) - 3.1\cos(x)\cos(y) + 5.2\sin(x)\cos(z) - 0.7 = 0$} & {\footnotesize $-0.1\cos(2z) - 4.1\cos(x)\cos(y) + 4.3\sin(x)\cos(z) - 2.2 = 0$} \\
    \midrule
    {\footnotesize (\romannumeral 2)} & {\footnotesize $-0.1\sin(2z) + 0.6\cos(y)\cos(z) + 4.9\sin(x)\sin(z) + 0.2 = 0$} & {\footnotesize $-0.9\sin(2z) + 2.4\cos(y)\cos(z) + 3.9\sin(x)\sin(z) + 0.2 = 0$} \\
    \midrule
    {\footnotesize (\romannumeral 3)} & {\footnotesize $-1.4\cos(x)\cos(z) + 3.7\cos(y)\sin(z) - 5.3\sin(x)\sin(y) + 0.3 = 0$} & {\footnotesize $+1.4\cos(x)\cos(z) - 3.7\cos(y)\sin(z) - 2.8\sin(x)\sin(y) + 0.6 = 0$} \\
    \midrule
    {\footnotesize (\romannumeral 4)} & {\footnotesize $-2.3\cos(x)\sin(z) + 5.2\sin(y)\cos(z) - 3.6\cos^2(x) + 0.4 = 0$} & {\footnotesize $-2.6\sin(x)\cos(y) + 5.2\sin(y)\cos(z) - 3.6\sin(x)\sin(y)\sin(z) - 0.3 = 0$} \\
    \midrule
    {\footnotesize (\romannumeral 5)} & {\footnotesize $-2.7\cos(y) - 3.9\cos(x)\cos(y) + 0.9\sin(x)\sin(z) - 0.6 = 0$} & {\footnotesize $-2.7\sin(2y) - 3.9\cos(x)\cos(y) + 2.9\sin(x)\sin(z) - 0.2 = 0$} \\
    \midrule
    {\footnotesize (\romannumeral 6)} & {\footnotesize $-3.1\cos(x)\cos(y) + 5.0\sin(y)\cos(z) + 3.3\sin(x)\sin(z) + 1.0 = 0$} & {\footnotesize $-3.7\cos(x)\cos(y) - 5.1\sin(y)\cos(z) - 2.3\sin(x)\sin(z) + 1.0 = 0$} \\
    \midrule
    {\footnotesize (\romannumeral 7)} & {\footnotesize $-4.8\cos(x)\sin(y) + 2.8\sin(x)\cos(z) + 1.7\sin^2(z) + 0.2 = 0$} & {\footnotesize $+4.8\cos(y)\sin(z) + 2.6\sin(x)\cos(z) - 1.6\sin^2(z) + 0.2 = 0$} \\
    \midrule
    {\footnotesize (\romannumeral 8)} & {\footnotesize $-5.1\cos(x) + 2.0\cos(z) + 4.1\cos(y)\cos(z) + 0.1 = 0$} & {\footnotesize $-5.1\cos(x) + 3.5\cos(z) - 4.6\cos(y)\cos(z) + 2.1 = 0$} \\
    \midrule
    {\footnotesize (\romannumeral 9)} & {\footnotesize $+0.3\cos(2z) - 2.3\cos(x)\cos(z) + 2.0\sin(x)\cos(y) + 0.2 = 0$} & {\footnotesize $+1.2\cos(2z) - 3.3\cos(x)\cos(z) + 4.4\sin(x)\cos(y) + 0.2 = 0$} \\
    \midrule
    {\footnotesize (\romannumeral 10)} & {\footnotesize $+0.5\cos(2x) - 5.4\cos(y)\sin(z) - 4.5\sin(x)\cos(y) - 0.2 = 0$} & {\footnotesize $+2.8\cos(2y) - 1.4\cos(y)\sin(z) - 4.5\sin(x)\sin(z) - 0.2 = 0$} \\
    \bottomrule
    \end{tabular}
    \captionsetup{justification=justified}
    \label{tab:SI-unconditional-novelty}
\end{table}

{
To further assess the effectiveness of DiffuMeta in generating valid shell geometries, we compare the 74\% validity rate achieved through unconditional generation with the success rate of randomly sampling geometries from the design space. Specifically, we randomly sample implicit equations by selecting random combinations of basis functions with random coefficients, following the same construction rules used to generate the training dataset as described in Section~\ref{sec:SI-dataset-generation}. Table~\ref{tab:random_sampling} presents the results of this random sampling approach across different sample sizes. As observed from Table~\ref{tab:random_sampling}, the random sampling approach yields an average validity rate of only $\sim$3.2\%, which is dramatically lower than the 74\% validity rate achieved by our approach. This stark contrast highlights the model's ability to learn and navigate the highly constrained and discontinuous design space of valid implicit surfaces.

\begin{table}[!ht]
    \centering
    \caption{{Validity rates for randomly sampled geometries from the design space.}}
    \label{tab:random_sampling}
    \begin{tabular}{ccccc}
    \toprule
    Total Samples & Valid Geometries & Invalid Geometries & Success Rate (\%) \\
    \midrule
    1,000 & 37 & 963 & 3.7 \\
    2,000 & 68 & 1,932 & 3.4 \\
    3,000 & 80 & 2,920 & 2.7 \\
    5,000 & 145 & 4,855 & 2.9 \\
    \midrule
    \multicolumn{3}{r}{\textbf{Average:}}& \textbf{3.2} \\
    \bottomrule
    \end{tabular}
\end{table}
    
    This extremely low success rate of random sampling reflects the sensitivity of triply periodic shell geometries to equation parameters: even a minor perturbation can drastically alter the resulting surface topology (or even lead to invalid structures, as illustrated in Fig.~S1(b)) and therefore their mechanical properties. Such sensitivity makes naive exploration of the design space infeasible, as there exists no analytical criterion to guarantee surface continuity or physical realizability from a given implicit equation alone, and, moreover, it is highly nontrivial to predict or design the mechanical properties from the analytical form. This reflects the complex, high-dimensional nature of the implicit design space. In contrast, our diffusion-based generative framework effectively learns the intricate correlations between equation parameters and valid geometric realizations directly from the equation sequences, enabling robust generation of continuous, connected, and mechanically meaningful structures. The resulting 74\% validity rate, therefore, demonstrates the framework’s capability to systematically explore the intricate design landscape of shell metamaterials far more effectively than random sampling.
}

\subsection{Implicit equations of conditionally generated shell designs}
\label{sec:SI-generated-equations}

Table~\ref{tab:SI-generated-equations} provides representative examples of the implicit surface equations for the shell metamaterial designs generated by DiffuMeta and shown in Figure 3 of the main manuscript. These results reveal several key insights into the complex relationships between implicit equations and mechanical properties. First, the model learns to utilize various trigonometric function combinations, enabling exploration of a broader design space beyond the classical minimal surface equations ({e.g.}, Schwarz P and D, Gyroids). Second, equations that appear mathematically similar can lead to distinct geometries (e.g., generated designs a(\romannumeral 2) and a(\romannumeral 4)), while seemingly disparate equations may produce similar mechanical responses through different deformation mechanisms (e.g., generated designs b(\romannumeral 1) and b(\romannumeral 2)). These observations highlight the complex mapping between implicit equations and the properties of resulting designs, where their physical behavior cannot be directly inferred from mathematical formulations alone. This complexity underscores the effectiveness of DiffuMeta's sequence-based approach in capturing and leveraging the subtle but critical relationships between implicit equation patterns and resulting metamaterial properties.

\begin{table}[htbp]
    \centering
    \caption{\textbf{Implicit surface equations for shell designs with target stress-strain responses (corresponding to Fig.~3 in the main article).} Each equation represents the mathematical description $\Psi(x,y,z) = 0$ that defines the corresponding shell topology. The diverse mathematical forms demonstrate the complex structure-property relationships in metamaterial design, where both geometrical features and mechanical properties cannot be readily predicted from the implicit equation formulations alone. Notably, generated designs a(\romannumeral 1) and b(\romannumeral 1) exhibit remarkably similar geometries despite distinct mechanical behaviors, while designs with vastly different mathematical expressions can produce comparable stress-strain responses.}
    \scalebox{0.95}{
    \begin{tabular}{clc}
    \toprule
    \textbf{Design} & \textbf{Implicit equation }$\Psi(x,y,z) = 0$ & \textbf{NRMSE (\%)} \\
    \midrule
    \multicolumn{3}{c}{\textbf{Target (a): Pronounced softening}} \\
    \midrule
    Best match & $2.5\cos(y)\cos(z) + 4.4\sin(x)\sin(y)\sin(z) - 0.6 = 0$ & 5.0 \\
    \midrule
    Inv. pred. (\romannumeral 1) & $3.2\cos(y)\sin(z) - 4.5\sin(x)\cos(z) + 2.1\sin^2(z) - 1.5 = 0$ & 3.3 \\
    \midrule
    Inv. pred. (\romannumeral 2) & $5.4\cos(y)\cos(z) - 3.9\sin(x)\sin(z) - 2.7\sin^2(z) + 1.9 = 0$ & 3.5 \\
    \midrule
    Inv. pred. (\romannumeral 3) & $4.2\cos(x)\cos(z) - 5.5\sin(y)\sin(z) - 2.6\sin^2(z) + 2.9 = 0$ & 3.6 \\
    \midrule
    Inv. pred. (\romannumeral 4) & $5.9\cos(y)\cos(z) - 5.0\sin(x)\sin(z) + 2.7\sin^2(z) - 2.9 = 0$ & 3.6 \\
    \midrule
    \multicolumn{3}{c}{\textbf{Target (b): Softening followed by hardening}} \\
    \midrule
    Best match & $-5.0\cos(x)\cos(y) + 5.9\sin(x)\sin(z) + 1.6\cos^2(x) + 2.4 = 0$ & 7.7 \\
    \midrule
    Inv. pred. (\romannumeral 1) & $-5.7\cos(y)\sin(z) + 3.9\sin(x)\cos(z) + 1.3 = 0$ & 4.6 \\
    \midrule
    Inv. pred. (\romannumeral 2) & $-5.0\cos(x)\sin(z) - 3.7\cos(y)\cos(z) + 1.2\sin(y)\cos(z) + 1.3 = 0$ & 7.4 \\
    \midrule
    Inv. pred. (\romannumeral 3) & $4.4\cos(y)\sin(z) + 3.4\sin(x)\cos(z) + 1.2 = 0$ & 8.4 \\
    \midrule
    Inv. pred. (\romannumeral 4) & $4.6\sin(y)\sin(z) + 3.2\sin(x)\cos(z) - 1.2 = 0$ & 8.5 \\
    \bottomrule
    \end{tabular}
    }
    \captionsetup{justification=justified}
    \label{tab:SI-generated-equations}
\end{table}

\begin{table}[htbp]
\centering
\caption{{\textbf{Implicit surface equations for the generated designs with multiple target properties (corresponding to Fig.~5 in the main article).}} The table includes the closest match from the dataset and the inverse predictions by {DiffuMeta}.}
\label{tab:SI-inverse-predictions-Fig5}
\begin{tabular}{clc}
\toprule
\textbf{Design} & \textbf{Implicit equation }$\Psi(x,y,z) = 0$ & \textbf{NRMSE (\%)}\\
\midrule
\multicolumn{3}{c}{\textbf{Multi-target conditional generation target (a)}} \\
\midrule
Best Match & $-2.4\sin(y)\cos(z) - 3.5\sin(x)\sin(z) + 5.5\sin^2(x) - 1.6 = 0$ & 11.2\\
\midrule
Inv. Pred. (\romannumeral 1) & $-3.9\sin(x) + 5.0\cos(2y) - 3.7\cos(y)\sin(z) + 1.1 = 0$ & 5.2\\
\midrule
Inv. Pred. (\romannumeral 2) & $5.4\sin(2y) - 5.7\cos(x)\cos(y) - 5.3\sin(y)\sin(z) + 0.9 = 0$ & 10.4\\
\midrule
Inv. Pred. (\romannumeral 3) & $-0.7\sin(2x) - 4.1\cos(y)\sin(z) + 0.2 = 0$ & 4.7\\
\midrule
\multicolumn{3}{c}{\textbf{Multi-target conditional generation target (b)}} \\
\midrule
Best Match & $-4.8\cos(y)\sin(z) - 0.9\sin(x)\cos(z) + 0.4 = 0$ & 22.1\\
\midrule
Inv. Pred. (\romannumeral 1) & $1.2\cos(x)\cos(z) + 3.9\cos(y)\sin(z) + 4.3\cos^2(y) - 0.6 = 0$ & 2.2\\
\midrule
Inv. Pred. (\romannumeral 2) & $-2.2\cos(2y) + 3.9\cos(y)\sin(z) + 1.3\sin(x)\cos(z) - 1.5 = 0$ & 3.5\\
\midrule
Inv. Pred.  (\romannumeral 3) & $-1.2\cos(x)\cos(z) + 3.9\cos(y)\sin(z) + 4.3\cos^2(y) - 0.6 = 0$ & 3.5\\
\bottomrule
\end{tabular}
\end{table}
\newpage
{
Table~\ref{tab:SI-inverse-predictions-Fig5} provides the implicit equations that correspond to the shell metamaterial designs generated by {DiffuMeta} and shown in Fig.~5 of the main article. We observe that all predicted equations remain {strictly within the defined design space bounds}, i.e., coefficient values within the range $(-6,6)$ with $0.1$ increments and the same trigonometric function basis. The apparent property deviations of some inverse predictions from the dataset in Fig.~5, therefore, do not stem from out-of-distribution coefficients, but rather from the model's ability to explore previously unseen regions of the property space. The key insight is that these predicted equations represent \emph{interpolation in the design space} (equation parameters) while achieving \emph{extrapolation in the property space} (mechanical behaviors). This capability highlights the strength of the generative modeling framework: it captures the {underlying structure-property relationships} governing complex, highly nonlinear mechanical behaviors, including effects of plasticity, contact, buckling, etc., all of which are highly nontrivial to capture or navigate using conventional optimization-based methods. This demonstrates the model's ability to generalize beyond the sampled data, effectively exploring new regions of the mechanical property landscape, which represents a key advantage of learned generative models over conventional search-based or optimization-based methods.

}

\subsection{Conditional generation on unseen target properties}

\begin{figure}[!htbp]
    \centering
    \includegraphics[width = \textwidth]{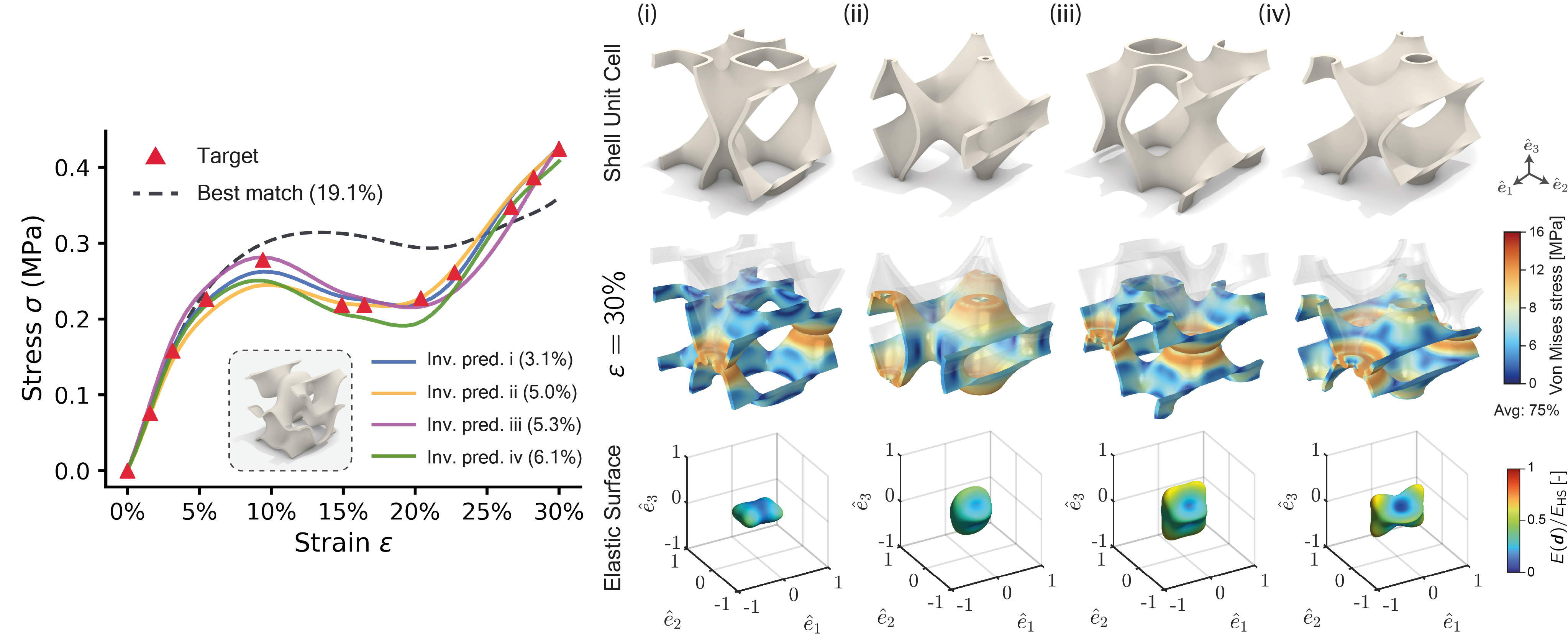}
    \captionsetup{justification=justified}
    \caption{\textbf{Inverse-designed shell metamaterials with extrapolated stress-strain responses}. The diffusion model generates multiple distinct designs conditioned on the target stress-strain curve (denoted by red triangles), which significantly extends beyond the training distribution (the best match from the training set is shown as well). The simulated deformed unit cells at $30\%$ strain are shown for both the best match from the training dataset and the generated designs. Elastic surface plots demonstrate distinct anisotropic elastic properties for each generated design. All shown stress-strain responses and elastic surfaces were obtained via finite element analysis, with NRMSE values (in brackets) quantifying design accuracy relative to target responses.}
    \label{fig:SI-extrapolation}
\end{figure}

In the main article, we present the results of the DiffuMeta model on a target response that lies outside the training distribution. Here, we examine the generalization capabilities of our model in generating multiple designs for unseen target responses. As shown in Fig.~\ref{fig:SI-extrapolation}, the model generates multiple designs that closely approximate the target response, achieving significantly lower errors ($3.1\%$–$6.1\%$) than the best match from the training dataset (NRMSE = $19.1\%$). 
The right panels in Fig.~\ref{fig:SI-extrapolation} further highlight the diversity and physical validity of the generated shells, demonstrating the model’s versatility to discover multiple, structurally unique designs for a given target, through different deformation and stress localization mechanisms, as shown in the von Mises stress distributions at $30\%$ compressive strain. Moreover, the elastic surface plots illustrate that the generated designs exhibit distinct anisotropic elastic properties, providing valuable design flexibility for multi-objective optimization scenarios. 

{
    While these results demonstrate successful extrapolation beyond the training distribution, it is important to acknowledge that, as a data-driven generative model, {DiffuMeta} learns to represent and interpolate patterns present in the training data, rather than to infer physical mechanisms beyond those expressible within the design space. In this work, the training data capture the mechanical responses of shell metamaterials, which, as observed in representative examples shown in Fig.~1c of the main article, tend to exhibit relatively smooth nonlinear stress-strain curves with progressive hardening or softening (such as the example in Fig.~5a exhibiting significant softening). Consequently, the model performs reliably for smooth nonlinear stress-strain responses, but its performance degrades when targeting mechanical behaviors that lie significantly beyond this learned distribution, such as responses with abrupt stress drops, sharp discontinuities, or highly oscillatory responses with multiple local extrema. When targeting responses outside this design space, the model tends to converge to the {closest physically plausible response} within the learned manifold, potentially approximating the target with smoother, simplified features or, in extreme cases, failing to identify valid designs altogether.

This behavior reflects a {fundamental property of generative models}: they favor statistically consistent and physically realizable regions of the data manifold. Therefore, to systematically assess the model's capabilities and limitations, we evaluate its performance by comparing the generated designs against the \emph{property boundaries} of the training dataset. As demonstrated in Figs.~4 and 5 of the main manuscript, the model successfully generates designs that {extend beyond existing property boundaries} (e.g., achieving more negative Poisson's ratios or previously unseen stress-strain responses). This demonstrates the model's ability to generalize and extrapolate within the learned physical constraints, rather than merely memorizing or interpolating between training examples. However, this extrapolation capability is inherently bounded by the underlying physics encoded in the training data. Targets that require fundamentally different mechanical mechanisms -- such as those arising from qualitatively distinct microstructural features or deformation modes not represented in the dataset -- lie beyond the model's generalization capacity. Improving performance on such targets would require expanding the training dataset to include {richer and more diverse mechanical responses}. For instance, incorporating structures with more complex internal architectures (e.g., hierarchical or graded designs), exploring broader ranges of geometric parameters, or including materials with more intricate constitutive behaviors (e.g., multi-stage plasticity, snap-through instabilities) could enable the model to learn and generate designs with correspondingly complex stress-strain characteristics, which could form a natural next step toward broader applicability of the framework.  

}

{
\subsection{Performance distribution of generated designs}

To assess the model's generation performance, we evaluated the distribution of generated design properties under both conditional and unconditional generation settings. First, to evaluate the model's reliability under conditional generation, we selected the target stress-strain responses shown in Fig.~3(a-b) of the main article and generated 100 candidate designs for each target. We note that the number of valid unique designs varied between the two cases (98 for target (a) vs.\ 38 for target (b)), reflecting differences in the inherent difficulty of the inverse design tasks. Each generated design was then analyzed through finite element simulations, and its mechanical response was compared to the target by computing the NRMSE values between the corresponding stress–strain curves.

\begin{figure}[h]
    \centering
    \includegraphics[width=0.82\textwidth]{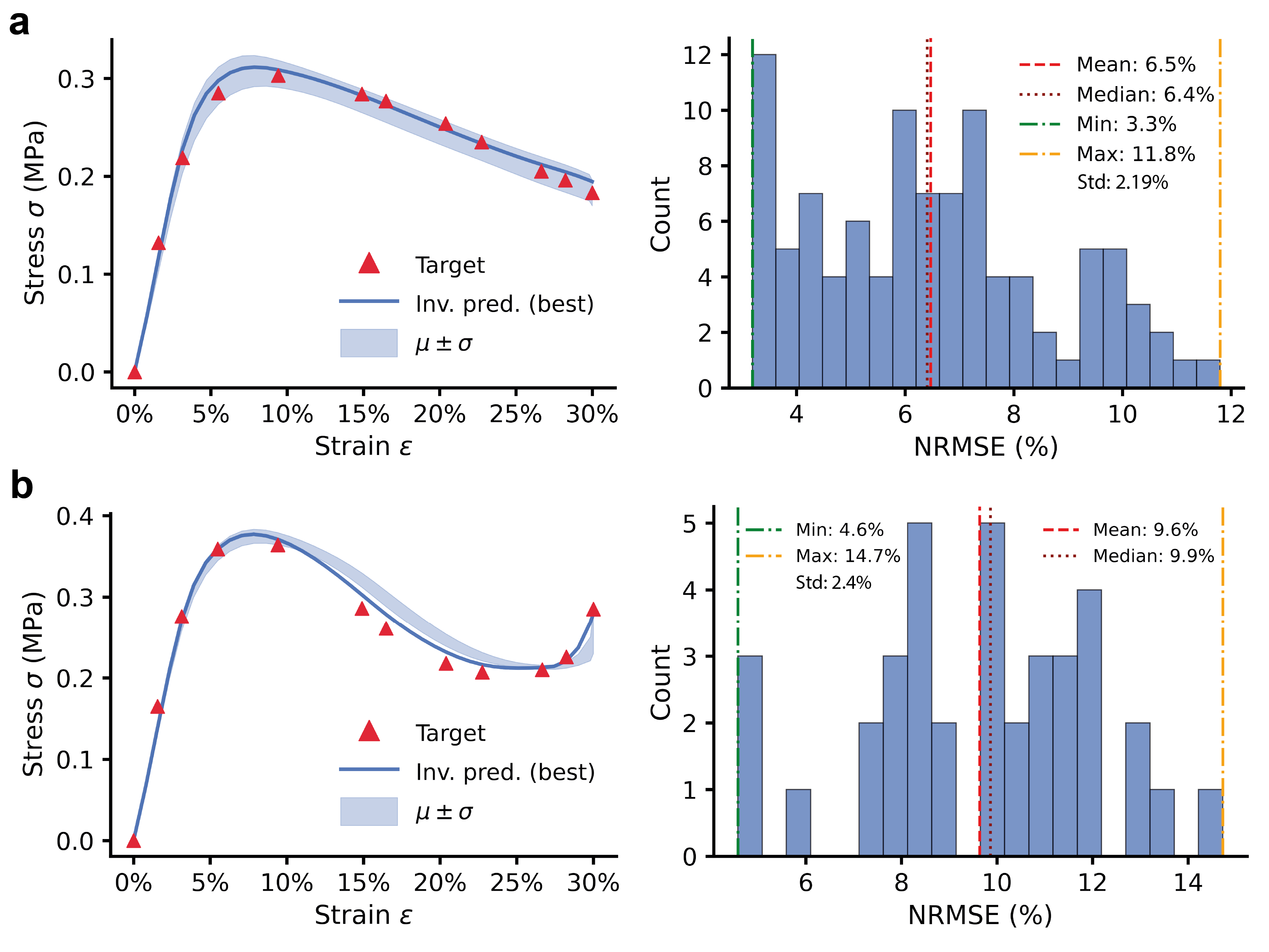}
    \caption{{\textbf{Performance distribution for conditional inverse design (corresponding to Fig.~3a--b in the main article).} The stress-strain plots show the target response (red triangles), best inverse prediction (solid blue line), and the mean $\pm$ one standard deviation interval (shaded blue region) across all valid generated designs. Also shown are the histogram plots of NRMSE values for generated designs. NRMSE values for target \textbf{(a)}: mean = 6.5\%, median = 6.4\%, best = 3.3\%, with 98 valid unique designs; for target \textbf{(b)}: mean = 9.6\%, median = 9.9\%, best = 4.6\%, with 38 valid unique designs.}}
    \label{fig:conditionally_generated_responses}
\end{figure}

As shown in Fig.~\ref{fig:conditionally_generated_responses}, the performance distributions reveal that the majority of generated designs consistently produce designs whose properties closely match the target responses with relatively small standard deviations, demonstrating stable and reliable conditional generation behavior. Importantly, the best-performing designs for both targets achieve NRMSE below 5\%, indicating good agreement with the prescribed properties. The slightly higher NRMSE values for target (b) suggest that some target responses are inherently more challenging to reproduce, likely due to their relative distance from the training data distribution or the complexity of the required deformation mechanisms (such as the hardening induced by contact among internal surfaces). Nevertheless, even for challenging targets, the model reliably produces designs with reasonable property matching at minimal computational cost ($\sim$0.2 seconds per design), offering efficient and versatile design exploration capabilities for practical applications.

\begin{figure}[!htp]
    \centering
    \includegraphics[width = \textwidth]{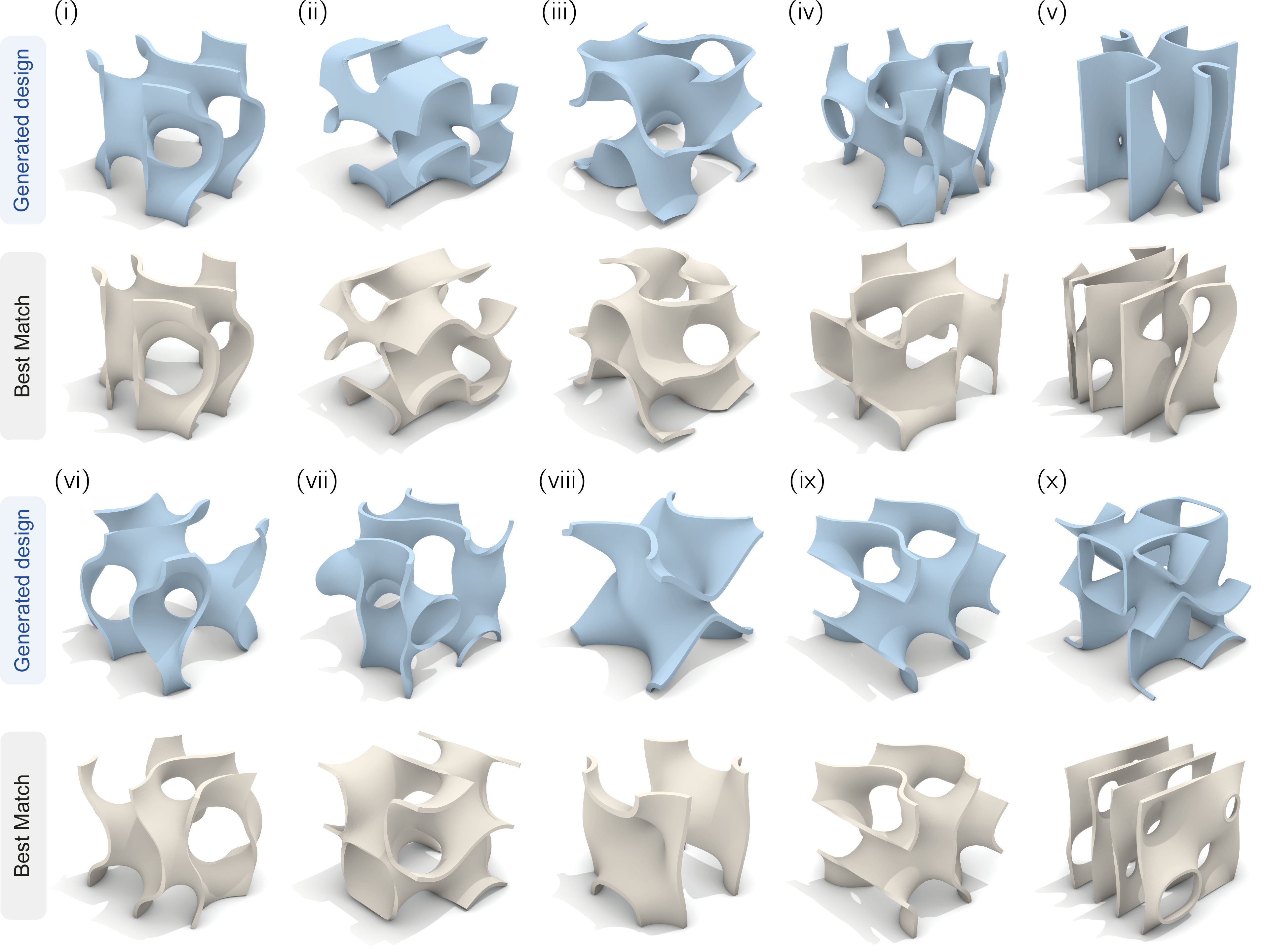}
    \captionsetup{justification=justified}
    \caption{\textbf{Representative examples of unconditionally generated shell metamaterial designs.} Visualizations of shell topologies generated by DiffuMeta through unconditional sampling and their most similar counterparts from the training dataset, corresponding to the equations listed in Table~\ref{tab:SI-unconditional-novelty}. Each design demonstrates unique geometric features, showcasing the model's ability to generate diverse metamaterial architectures with novel topologies.}
    \label{fig:SI-unconditional-novelty}
\end{figure}

\begin{figure}[!ht]
    \centering
    \includegraphics[width=\textwidth]{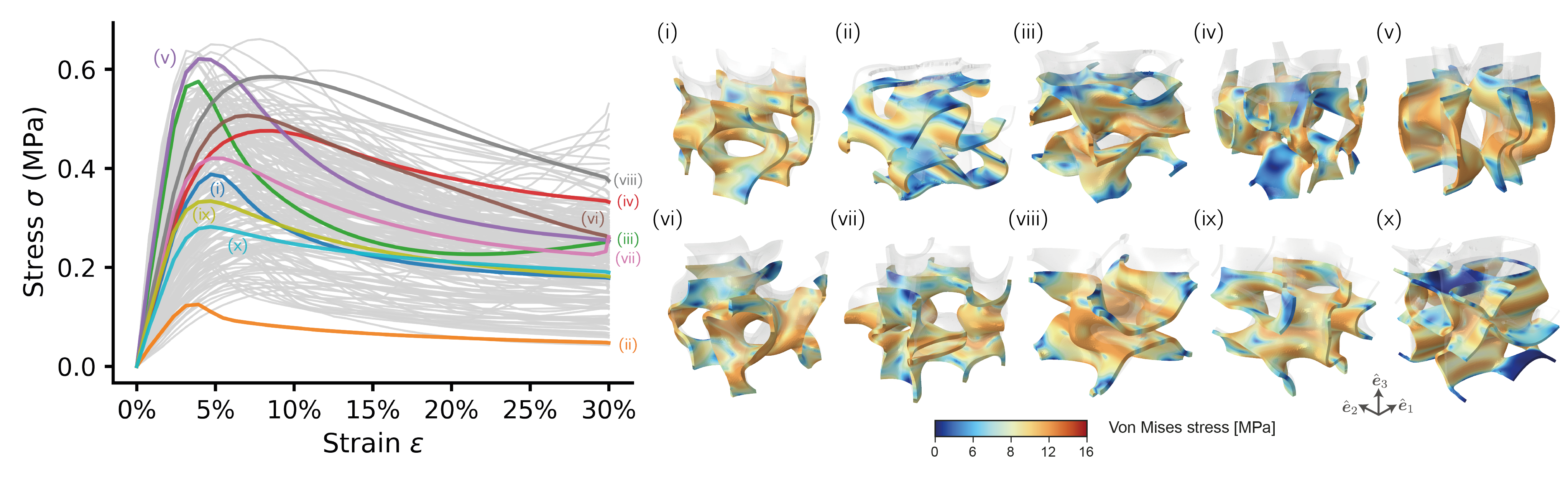}
    \caption{{\textbf{Stress-strain responses of unconditionally generated designs.} Stress-strain responses for 10 representative examples of unconditionally generated shell metamaterial designs (colored curves) compared with the stress-strain responses of 200 structures randomly sampled from the training dataset shown as reference (gray curves). The generated designs span a wide range of mechanical behaviors across the property space. Also shown are the corresponding deformed unit cells for 10 representative designs (\romannumeral 1--\romannumeral 10), colored by the von Mises stress distribution at 30\% compressive strain, demonstrating significant topological and geometric diversity.}}
    \label{fig:SI-unconditionally-generated-responses}
\end{figure}

To further validate that the model has captured the full diversity of the design space rather than merely memorizing training examples, we evaluated unconditional generation performance. We selected 10 representative examples generated under unconditional sampling and computed their stress-strain responses through FE simulation (corresponding to Table~\ref{tab:SI-unconditional-novelty} and Fig.~\ref{fig:SI-unconditional-novelty}). As shown in Fig.~\ref{fig:SI-unconditionally-generated-responses}, the unconditionally generated designs (highlighted curves) span a wide range of mechanical behaviors that are well-distributed across the property space covered by the training dataset (gray curves). The highlighted representative designs (\romannumeral 1--\romannumeral 10) demonstrate remarkable geometric diversity, exhibiting distinct shell topologies with varying curvature patterns and local features. For each representative case, we also identify the closest counterpart from the training set based on similarity in their implicit equations, as shown in Table~\ref{tab:SI-unconditional-novelty} and Fig.~\ref{fig:SI-unconditional-novelty}. This diversity confirms that the model has effectively learned the underlying distribution of the design space and can generate novel, structurally valid metamaterial architectures with diverse mechanical responses, rather than simply reproducing known structures from the training set.

Together, these results demonstrate that the model achieves \emph{stable, reliable, and statistically consistent generation performance} for inverse design tasks, and more importantly, at negligible computational costs. This allows users to generate multiple candidates and select designs based on additional criteria such as manufacturability, geometric features, or robustness considerations, thereby leveraging the probabilistic nature of generative models to address the one-to-many mapping challenge in inverse design tasks.
}
\newpage
{
\subsection{Robustness and practical considerations of generated  designs}

We have demonstrated DiffuMeta's ability to generate diverse shell metamaterials with precisely targeted mechanical properties. Beyond this, an important consideration is the robustness and manufacturability of these designs in practical applications. While shell metamaterials can exhibit complex mechanical behaviors that are, in certain regimes, sensitive to geometric perturbations such as variations in surface thickness or local curvature, the shell architectures are inherently resilient to such effects during manufacturing and operation. In the following, we discuss several key factors that contribute to the robustness and practical applicability of the inverse-designed structures in practice.

First, our choice of implicit equation-based parameterization accurately preserves the details of the geometric features while avoiding discretization artifacts or imperfections that are common in explicit representations \cite{jadhav2024generative} (e.g., voxels or point clouds). This ensures high-fidelity reproduction of the designed geometry during manufacturing. Furthermore, shell-based metamaterials inherently avoid the stress concentrations that commonly affect beam-based and plate-based architectures at junctions of structural members \cite{portela2018impactb,latture2018effectsa,tancogne-dejean20183d}, which can lead to poor recoverability and potential failure. In addition, the continuous topology of shell lattices offers key advantages for advanced manufacturability. In contrast to closed-cell plate-based metamaterials that pose significant challenges in powder or liquid extraction during 3D printing \cite{tancogne-dejean20183d,meyer2024nonsymmetric}, our open-cell shell architectures facilitate efficient removal of unmolten powders or residual resins in powder-bed or liquid resin-based additive manufacturing processes \cite{wang2022achieving,ma2021elasticallyisotropic}. Their smooth, continuous surfaces also alleviate overhang constraints, ensuring high-quality, high-fidelity fabrication \cite{ma2025multiphysical,lu2022architectural}. (Besides, prior plate-based designs \cite{tancogne-dejean20183d} have shown that introducing punctured holes in the stretching-dominated elements--similar to the shells in our study--has little influence on the effective response.)

It is also important to note that the target applications most relevant to this work -- such as cushioning, padding, or energy absorption -- involve distributed loading conditions rather than concentrated loads \cite{ha2023rapidb,maurizi2025designing,dalaq2023origamiinspired}. For these use cases, the \textit{global stress-strain response} is the dominant design criterion, and the \textit{overall} performance is comparatively tolerant to small, localized defects or imperfections that may occur during use \cite{montanari2023defect,niutta2022influence}. Furthermore, to enhance the robustness of designed structures in application-critical scenarios, additional design constraints can be directly incorporated into our framework. For instance, the training dataset can be augmented with perturbed geometries (e.g., localized thickness variations, surface imperfections) and their corresponding properties, enabling the model to learn robust designs. Alternatively, robustness metrics such as performance degradation under geometric perturbations can be added as conditioning targets alongside mechanical properties. In addition, the framework could be extended to quantify and propagate uncertainty by integrating Bayesian inference into the diffusion model \cite{kou2024bayesdiff,jazbec2025generative}, thereby enabling estimation of confidence intervals for the generated designs and their predicted properties.
}
\bibliography{SI-Bib}